\documentclass[11pt,a4paper]{article}
\usepackage{jinstpub}

\usepackage{ifthen} 
\usepackage{subcaption}
\usepackage{graphicx}
\usepackage{amsmath}
\usepackage{bm}
\usepackage[dvipsnames]{xcolor}
\usepackage[range-phrase=--,%
    list-units=single,%
    range-units=single,%
    per-mode=reciprocal]{siunitx}
\usepackage{placeins}
\usepackage{wrapfig}

\usepackage[nottoc,numbib]{tocbibind}
\usepackage[noabbrev]{cleveref}
\usepackage[cachedir=minted/,%
    frozencache=true
    ]{minted}
\setminted[c++]{xleftmargin=\parindent, xrightmargin=\parindent, frame=single}

\newboolean{uprightparticles}
\setboolean{uprightparticles}{false}

\usepackage{xspace} 
\usepackage{upgreek}

\def\babar  {\mbox{BaBar}\xspace}

\def\MagUp {\mbox{\em Mag\kern -0.05em Up}\xspace}

\ifthenelse{\boolean{uprightparticles}}%
{

 \def\PDelta      {\ensuremath{\Delta}\xspace}                 
 \def\PXi      {\ensuremath{\Xi}\xspace}                 
 \def\PLambda      {\ensuremath{\Lambda}\xspace}                 
 \def\PSigma      {\ensuremath{\Sigma}\xspace}                 
 \def\POmega      {\ensuremath{\Omega}\xspace}                 
 \def\PUpsilon      {\ensuremath{\Upsilon}\xspace}                 
 
 \def\PB      {\ensuremath{\mathrm{B}}\xspace}                 
                  
 \def\PD      {\ensuremath{\mathrm{D}}\xspace}

 \def\PK      {\ensuremath{\mathrm{K}}\xspace}

 \def\Pi      {\ensuremath{\mathrm{i}}\xspace}

 \def\Ps      {\ensuremath{\mathrm{s}}\xspace}

}
{

 \mathchardef\PDelta="7101
 \mathchardef\PXi="7104
 \mathchardef\PLambda="7103
 \mathchardef\PSigma="7106
 \mathchardef\POmega="710A
 \mathchardef\PUpsilon="7107
                  
 \def\PB      {\ensuremath{B}\xspace}                 
                  
 \def\PD      {\ensuremath{D}\xspace}

 \def\PK      {\ensuremath{K}\xspace}

 \def\Pi      {\ensuremath{i}\xspace}

 \def\Ps      {\ensuremath{s}\xspace}

}

\makeatletter
\ifcase \@ptsize \relax
  \newcommand{\miniscule}{\@setfontsize\miniscule{4}{5}}
\or
  \newcommand{\miniscule}{\@setfontsize\miniscule{5}{6}}
\or
  \newcommand{\miniscule}{\@setfontsize\miniscule{5}{6}}
\fi
\makeatother

\DeclareRobustCommand{\optbar}[1]{\shortstack{{\miniscule (\rule[.5ex]{1.25em}{.18mm})}
  \\ [-.7ex] $#1$}}

\def\squark    {{\ensuremath{\Ps}}\xspace}

  \def\Kbar    {{\kern 0.2em\overline{\kern -0.2em \PK}{}}\xspace}

\def\KorKbar    {\kern 0.18em\optbar{\kern -0.18em K}{}\xspace}

  \def\Dbar    {{\kern 0.2em\overline{\kern -0.2em \PD}{}}\xspace}

\def\DorDbar    {\kern 0.18em\optbar{\kern -0.18em D}{}\xspace}

\def\B       {{\ensuremath{\PB}}\xspace}
\def\Bbar    {{\ensuremath{\kern 0.18em\overline{\kern -0.18em \PB}{}}}\xspace}

\def\BorBbar    {\kern 0.18em\optbar{\kern -0.18em B}{}\xspace}

\def\Bs      {{\ensuremath{\B^0_\squark}}\xspace}

  \def\Y#1S{\ensuremath{\PUpsilon{(#1S)}}\xspace}

\def\Lbar        {{\ensuremath{\kern 0.1em\overline{\kern -0.1em\PLambda}}}\xspace}
\def\LorLbar    {\kern 0.18em\optbar{\kern -0.18em \PLambda}{}\xspace}

\def\to                 {\ensuremath{\rightarrow}\xspace}

\def\AT#1     {\ensuremath{A_{\mathrm{T}}^{#1}}\xspace}           

\def\C#1      {\ensuremath{\mathcal{C}_{#1}}\xspace}                       
\def\Cp#1     {\ensuremath{\mathcal{C}_{#1}^{'}}\xspace}                    
\def\Ceff#1   {\ensuremath{\mathcal{C}_{#1}^{\mathrm{(eff)}}}\xspace}        
\def\Cpeff#1  {\ensuremath{\mathcal{C}_{#1}^{'\mathrm{(eff)}}}\xspace}       
\def\Ope#1    {\ensuremath{\mathcal{O}_{#1}}\xspace}                       
\def\Opep#1   {\ensuremath{\mathcal{O}_{#1}^{'}}\xspace}                    

\newcommand{\tev}{\ensuremath{\mathrm{\,Te\kern -0.1em V}}\xspace}
\newcommand{\gev}{\ensuremath{\mathrm{\,Ge\kern -0.1em V}}\xspace}
\newcommand{\mev}{\ensuremath{\mathrm{\,Me\kern -0.1em V}}\xspace}
\newcommand{\kev}{\ensuremath{\mathrm{\,ke\kern -0.1em V}}\xspace}
\newcommand{\ev}{\ensuremath{\mathrm{\,e\kern -0.1em V}}\xspace}
\newcommand{\gevc}{\ensuremath{{\mathrm{\,Ge\kern -0.1em V\!/}c}}\xspace}
\newcommand{\mevc}{\ensuremath{{\mathrm{\,Me\kern -0.1em V\!/}c}}\xspace}
\newcommand{\gevcc}{\ensuremath{{\mathrm{\,Ge\kern -0.1em V\!/}c^2}}\xspace}
\newcommand{\gevgevcccc}{\ensuremath{{\mathrm{\,Ge\kern -0.1em V^2\!/}c^4}}\xspace}
\newcommand{\mevcc}{\ensuremath{{\mathrm{\,Me\kern -0.1em V\!/}c^2}}\xspace}

\def\gsim{{~\raise.15em\hbox{$>$}\kern-.85em
          \lower.35em\hbox{$\sim$}~}\xspace}
\def\lsim{{~\raise.15em\hbox{$<$}\kern-.85em
          \lower.35em\hbox{$\sim$}~}\xspace}

\def\roofit     {\mbox{\textsc{RooFit}}\xspace}

\def\root       {\mbox{\textsc{Root}}\xspace}

\def\tell1  {TELL1\xspace}
\def\ukl1   {UKL1\xspace}

\def\tt{\ensuremath{\tau\tau}\xspace}

\def\Nev{\ensuremath{N_{\text{ev}}}\xspace}
\def\Nbins{\ensuremath{N_{\text{bins}}}\xspace}
\def\fsig{\ensuremath{f_{\text{sig}}}\xspace}
\def\PDFs {PDFs\xspace}

\title{Avoiding biases in binned fits}

\abstract{
Binned maximum likelihood fits are an attractive option when analysing large datasets, but require care when computing likelihoods of continuous PDFs in bins.
For many years the widely used  statistical modelling package \roofit evaluated probabilities at the bin centre, leading to significant biases for strongly curved probability density functions.
We demonstrate the biases with real-world examples, and introduce a PDF class to \roofit that removes these biases.
The physics and computation performance of this new class are discussed.
}

\author[a]{V.~V.~Gligorov,}
\author[b,1]{S.~Hageboeck, \note{Corresponding author.}}
\author[c]{T.~Nanut,}
\author[d]{A.~Sciandra,}
\author[a]{and D.~Y.~Tou}

\affiliation[a]{LPNHE, \\
Sorbonne Université, Paris Diderot Sorbonne Paris Cité, CNRS/IN2P3}
\affiliation[b]{CERN,
1 Esplanade des Particules, Meyrin, Switzerland}
\affiliation[c]{Institute of Physics, Ecole Polytechnique F\'{e}d\'{e}rale de Lausanne (EPFL), Lausanne, Switzerland}
\affiliation[d]{Santa Cruz Institute of Particle Physics, University of California at Santa Cruz,
5 High Street, Santa Cruz, CA, USA}
\emailAdd{stephan.hageboeck@cern.ch}


\begin{document}

\maketitle

\section{Introduction}
\roofit~\cite{roofit,RooFitManual} is a C++ package for statistical modelling distributed with \root~\cite{ROOT,ROOTManual}. \roofit's development started in the year 2000 within the \babar collaboration.
Since then, \roofit has been a reliable tool for many experiments in high-energy physics (HEP) at $B$ factories and the Large Hadron Collider.
With \roofit, users can define likelihood models using observables, parameters, functions and \PDFs \footnote{\texttt{P}robability \texttt{D}ensity \texttt{F}unctions}, which can be fitted to data, plotted or used for statistical tests.

\begin{wrapfigure}{r}{0.45\textwidth}
\sbox0{\includegraphics{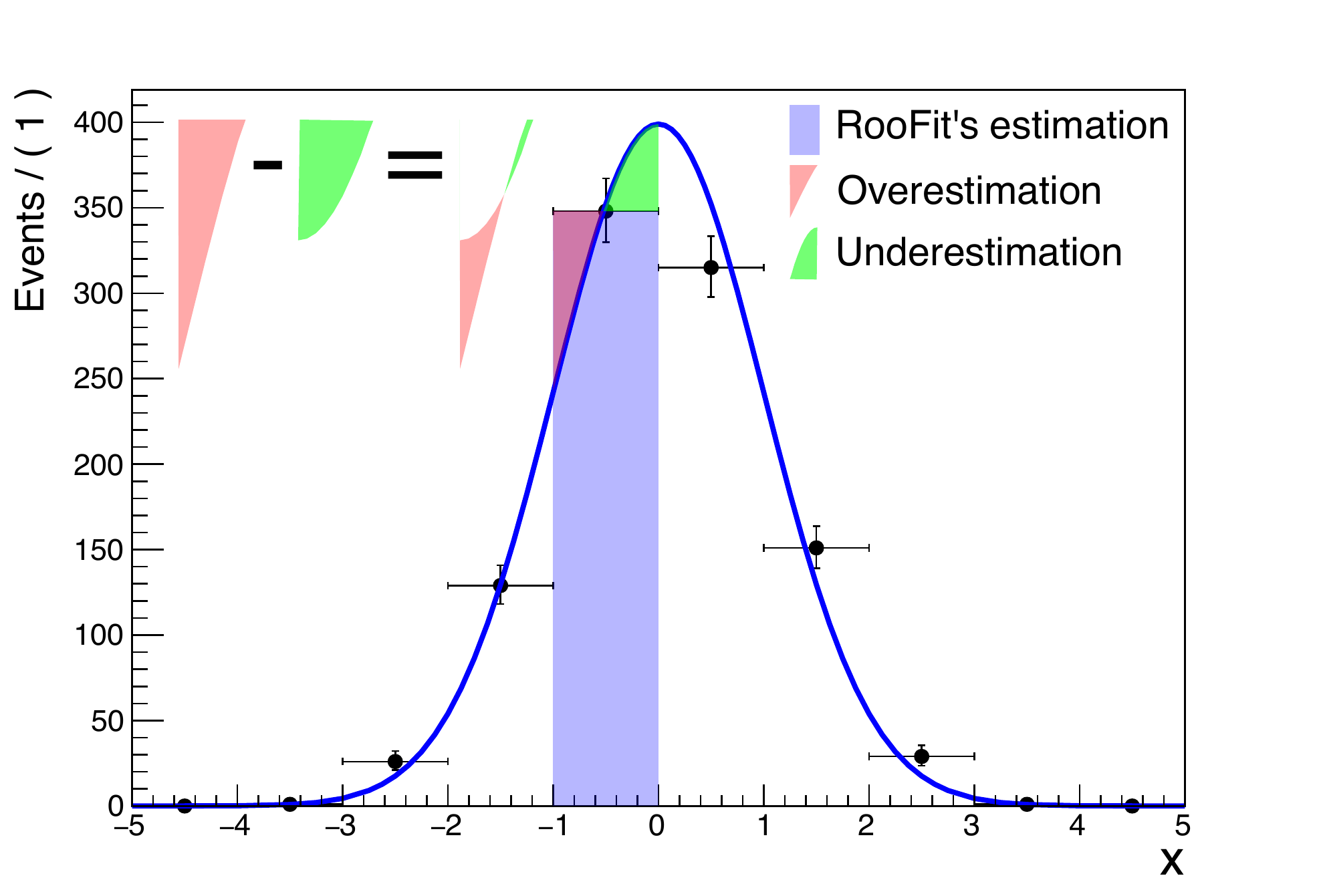}}%
\raggedleft
\includegraphics[clip=false,%
    trim={.05\wd0} {.1\ht0} {.09\wd0} {.12\ht0},%
    width=0.42\textwidth]{figures/binnedFitIllust_largeFonts.pdf}
\captionof{figure}{Error of estimating the probability in a bin by evaluating a PDF at the bin centre.}
\label{fig:binnedFitIllust}
\end{wrapfigure}

As HEP datasets become progressively larger, it becomes increasingly important to reduce the computation time of data analysis tasks such as fits.
At the same time, the higher statistics allow for measurements with higher precision.
These demands of speed and accuracy can come into conflict. In this paper, we address the long-standing problem of a bias in \roofit's binned likelihood fits, while maintaining much of their speed
advantage over unbinned likelihood fits.




These biases arise since in binned fits, \roofit samples probability densities at only one point in a bin, the bin centre.
It is assumed that this is a decent approximation for the probability of the entire bin.
When distributions are strongly curved, though, this is inaccurate.
\Cref{fig:binnedFitIllust} illustrates the error of this approximation with a Gaussian distribution.
For the shown bin, the probability to the left of the bin centre is overestimated by the surface shaded in red.
To the right of the bin centre, it is underestimated by the surface shaded in green.
If the PDF was linear in the bin, these two surfaces would have equal size.
However, if the PDF has a non-vanishing second derivative, the compensation is insufficient.
In the marked bin, the probability is  overestimated. 

Although well known, removing this bias was considered too invasive for \roofit.
Users were encouraged to use more bins to reduce biases, which is not always possible, though.
We present here the first rigorous solution, extending \roofit with a new PDF class, which transforms a continuous into a binned PDF.
This leaves likelihood and $\chi^2$ calculation functions, as well as plotting routines, unaffected, but eliminates the bias.
We demonstrate the biases with real-world examples, and discuss the physics and computation performance using this PDF class. The new PDF is available in \root since version 6.24.


\section{Integrating continuous PDFs for unbiased binned fits}
\label{sec:BinSamplingPdf}

To eliminate biases in binned fits, the PDF class ``\textsc{RooBinSamplingPdf}''~\cite{RooFit_ICHEP20,BinSamplingPdf} was added to \roofit.
It transforms continuous PDFs into binned PDFs by integrating the former in each bin and dividing the result by the bin width.
The resulting PDF is constant in each bin, and evaluates to the \textit{average} probability density in a bin (instead of the probability density at the bin centre).
This allows for fitting continuous PDFs to binned data, or for adding binned and continuous PDFs to create sum models.

To integrate the original PDF, \root's \textsc{IntegratorOneDim} is used.
Internally, this uses the adaptive Gauss-Kronrod~\cite{GaussKronrod} integrator with a 21-point rule from the GSL~\cite{gsl}.\footnote{In environments where \root is used without the GSL, \root's non-adaptive \textsc{GaussIntegrator} is used.}
For smooth functions, this means that 21 times more function evaluations are required compared to using a simple binned PDF.
If the adaptive algorithm has to subdivide single bins to improve the integration accuracy, the computational overhead increases to a bit more than 21-fold, but this happens only in bins where the original PDF is not smooth.
Nevertheless, since binned fits are usually orders of magnitudes faster than running an unbinned fit, see \cref{sec:timeComplexity}, the additional time to perform a more accurate integration is still acceptable if users get unbiased results.

Users can further set the relative precision required for the integration.
They can even directly manipulate all settings of \root's \textsc{IntegratorOneDim} to customise accuracy and speed, e.g.\ switch to a 15-point Gauss-Kronrod rule. 

\begin{figure}[t]
    \centering
    \begin{subfigure}[b]{0.45\textwidth}
        \centering
        \sbox0{\includegraphics{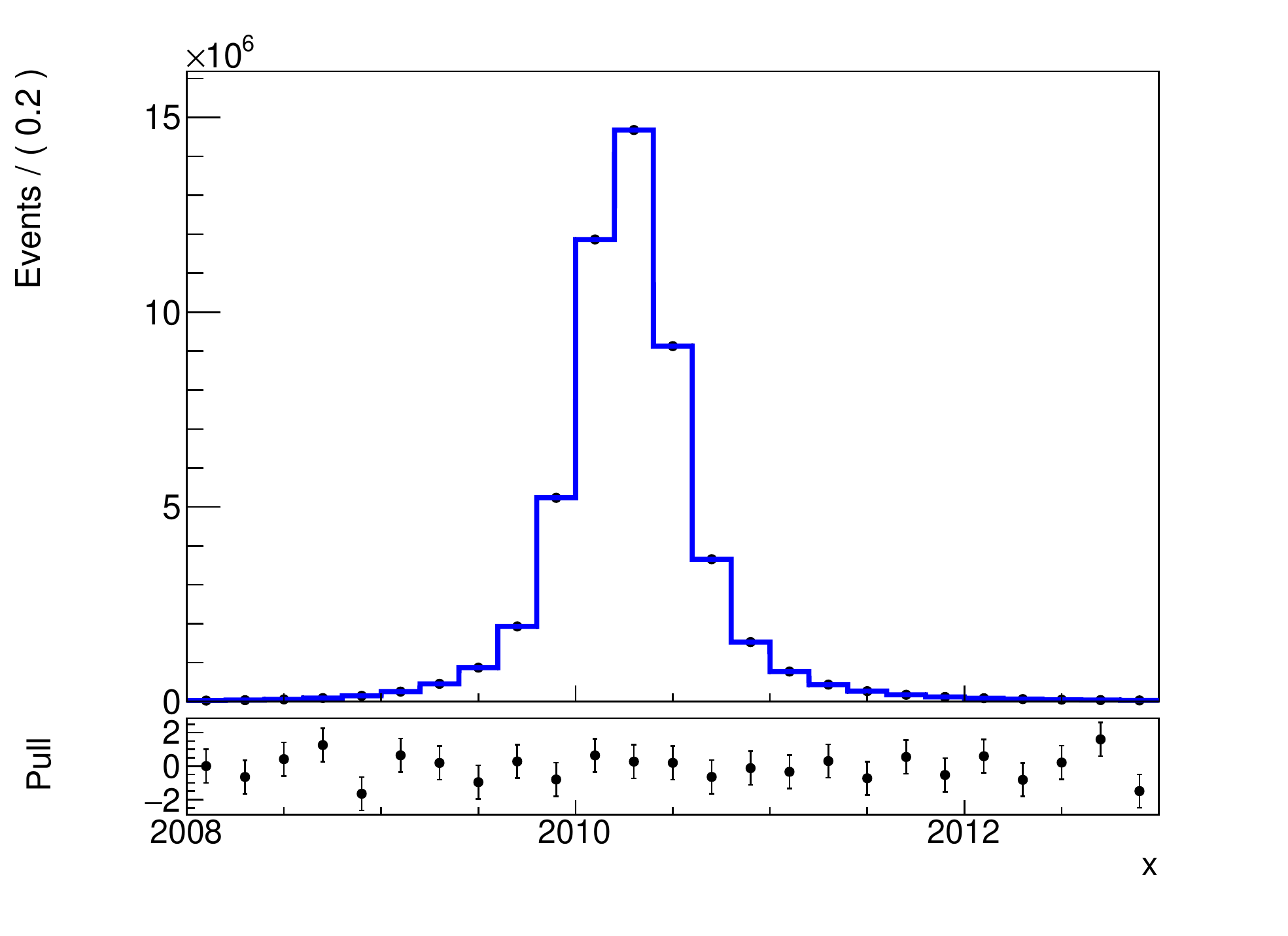}}%
        \includegraphics[clip=false,%
            trim={.05\wd0} {.1\ht0} {.1\wd0} {.05\ht0},%
            width=\textwidth]{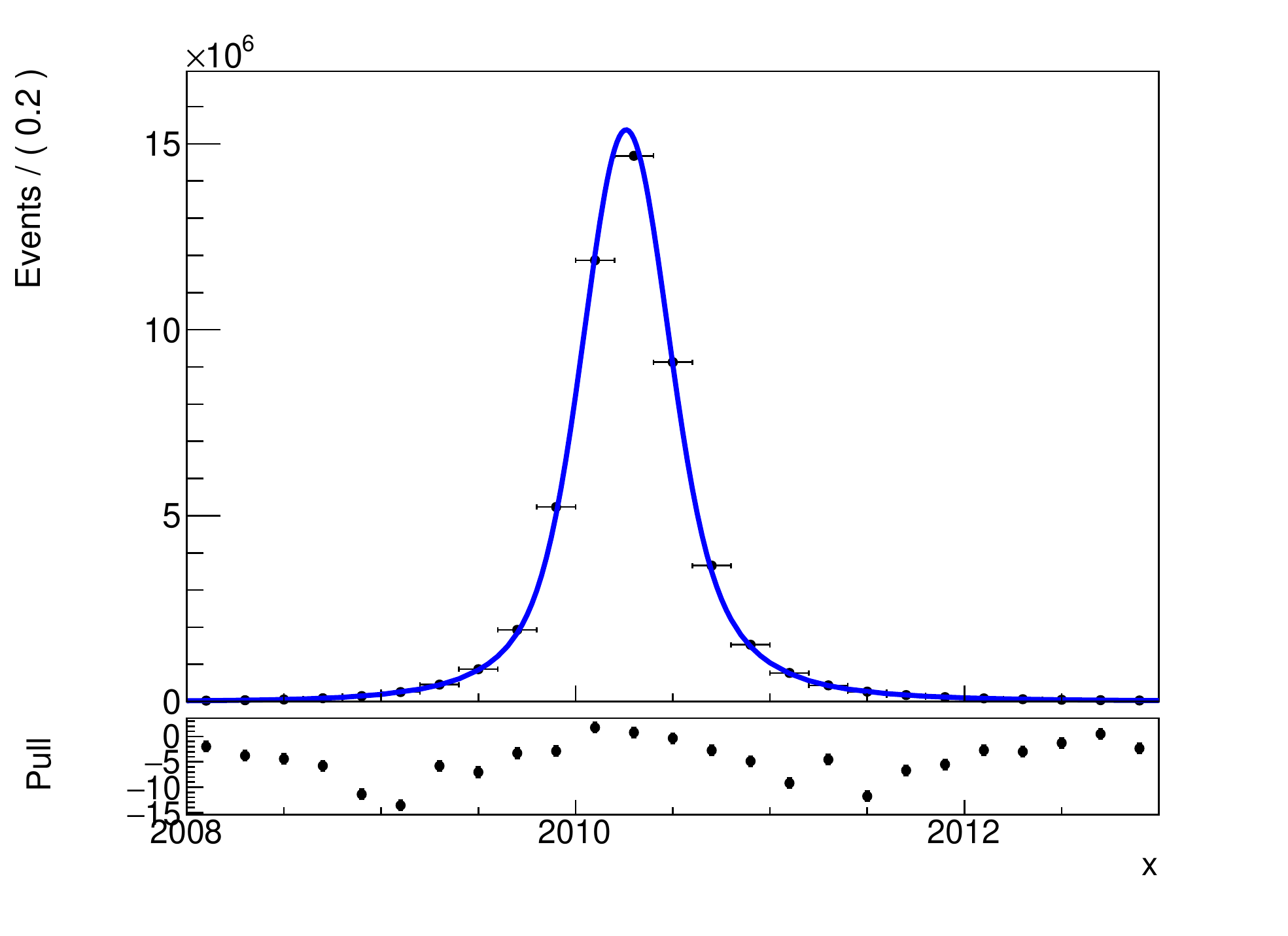}
        \caption{Continuous PDF compared to binned data}
        \label{fig:pulls_continuousPdf}
    \end{subfigure}
    \hfil
    \begin{subfigure}[b]{0.45\textwidth}
        \centering
        \sbox0{\includegraphics{figures/testCharm_binned.pdf}}%
        \includegraphics[clip=false,%
            trim={.05\wd0} {.1\ht0} {.1\wd0} {.05\ht0},%
            width=\textwidth]{figures/testCharm_binned}
        \caption{PDF wrapped into \textsc{RooBinSamplingPdf}}
        \label{fig:pulls_binSamplingPdf}
    \end{subfigure}
    \caption{Comparison of pulls without and with \textsc{RooBinSamplingPdf}~\cite{RooFit_ICHEP20}. The bottom pads show pulls that are computed by comparing data counts with the plotted curves evaluated at the bin centres. In (\subref{fig:pulls_binSamplingPdf}), the $y$-axis of the pull plot is zoomed 4-fold.}
    \label{fig:comparisonBinSamplingPdf}
\end{figure}

\Cref{fig:comparisonBinSamplingPdf} shows a fit model that is a sum of three Gaussian distributions and a JohnsonSU distribution~\cite{Johnson:1949zj} --- a typical signal model for analysing charm decays~\cite{CPVCharm}.
When \textsc{RooBinSamplingPdf} is used, the probability density represented by the PDF is constant in each bin, whereas it is strongly curved using the original model.
The pads at the bottom show pulls\footnote{``Pull'' denotes a residual divided by the standard deviation of a quantity. In this work, this term is used both to quantify a deviation of model parameters from a theoretical value as well as for comparing a fitted distribution with data.} comparing the plotted curves evaluated at the bin centres and data points sampled from the model.
Due to the over- or underestimation illustrated in \cref{fig:binnedFitIllust}, strong pulls are observed in \cref{fig:pulls_continuousPdf}.
Since a fitter will try to balance the overshoot in the tails of the signal model with the undershoot in the centre, fit results will be biased.
\Cref{fig:pulls_binSamplingPdf} shows that the use of \textsc{RooBinSamplingPdf} eliminates the pulls.

Users have two ways to use this new PDF class.
First, they can directly construct an instance of \textsc{RooBinSamplingPdf}.
They pass an observable that defines the desired binning, the continuous PDF, and (if desired) the relative precision for bin integrals:
\begin{minted}[fontsize=\footnotesize]{c++}
RooBinSamplingPdf binSampler("Name", "Plot Label",
                             binnedObservable, originalPDF,
                             /*optional integrator precision=*/ 1.E-4);
binSampler.fitTo(data);
\end{minted}

Second, \roofit can be instructed to automatically construct instances of \textsc{RooBinSamplingPdf} before each fit:
\begin{minted}[fontsize=\footnotesize,escapeinside=||]{c++}
pdf.fitTo(data, |\textcolor{BrickRed}{RooFit::IntegrateBins}|(<integrator precision>));
\end{minted}

The first strategy is most versatile, because users can control which PDFs are transformed, and which integrators are used.
Moreover, plots of \textsc{RooBinSamplingPdf} correctly convey the fact that a binned fit was used, and pull plots will also benefit from the correction (cf.\ \cref{fig:comparisonBinSamplingPdf}).

The second strategy is particularly useful in simultaneous fits with multiple channels, where a mixture of binned and unbinned fits is used.
When \mintinline{c++}{RooFit::IntegrateBins(0.)} is passed to the fit instruction, \roofit will apply the correction to all continuous PDFs that are fit to binned data, but leave unbinned fits unchanged.
Passing a non-zero value, the correction will be applied to all channels, irrespective of whether a binned or unbinned data set is used.
This is important since some fitting frameworks simulate binned fits by creating an ``unbinned'' data set, where the coordinates of an entry correspond to a bin centre, and the weight of an entry corresponds to the bin content.
Although this produces the same result as a binned fit, \roofit cannot detect that the correction discussed in this work needs to be applied, so users need a way to explicitly enable it.

Although the bias in binned fits could also have been corrected for by changing the likelihood calculation functions in \roofit, adding the new PDF class has advantages:
\begin{itemize}
    \item It is not invasive. No changes to \roofit's fitting algorithms are required. Since the probability density of \textsc{RooBinSamplingPdf} is constant in each bin, \roofit's default strategy of evaluating the function in the centre and multiplying by the bin width is correct even if the original PDF is curved strongly.
    \item It is reusable. The same strategy that is used to correct likelihood fits can be used to correct $\chi^2$ fits or the computation of $\chi^2$ statistics (cf.\ \cref{fig:LHCb-charm-chi2}, \cref{sec:lhcb_charm}).
    \item It allows for better plotting. Since data are usually plotted in bins, plotting a binned PDF allows for inspecting differences by eye. When plotting binned data and a continuous PDF, the viewer would have to estimate the integral of the PDF over a bin.
    Moreover, the new class overrides the function \mintinline{c++}{RooAbsArg::plotSamplingHint()}.
    This function conveys the optimal points for plotting a function to \roofit's plotting routines.
    If this function would not be overridden, for example since the correction was implemented in the likelihood functions, \roofit would employ heuristics to estimate the best vertices for drawing a curve, which can lead to binning artifacts in the plotted curves.
    \item It is achieved with a limited computational overhead, which depends on the integrator settings. Independent of the integration in bins, \roofit needs to ensure that all PDFs are normalised, which means that they need to be integrated over the fit range. If no analytic integral is known, numeric integration algorithms are employed, which would typically evaluate a PDF at \numrange{64}{256} points. If \textsc{RooBinSamplingPdf} is used, however, no further automatic normalisation needs to be performed, since the integrals in the bins can be reused to compute the integral over the fit range. The computational overhead therefore only depends on the integrator settings of \textsc{RooBinSamplingPdf}, but the integration step that would normally be executed automatically by \roofit is skipped.
\end{itemize}

\subsection{Time complexity of likelihood calculations}
\label{sec:timeComplexity}

An unbinned fit in \roofit estimates the parameters of a model by maximising the negative logarithmic likelihood, $\mathcal{L}$, defined as
\begin{equation}
-\log\mathcal{L} = - \sum^{N}_{i=1}\log{f(x_{i} \mathbin{|} \bm{\theta})},
\label{eq:binnedlikelihood}
\end{equation}
where $f(x_{i} \mathbin{|} \bm{\theta})$ is the PDF of the model, and the likelihood is a product over $N$ observations $x_{i}$. The term $f(x_{i} \mathbin{|} \bm{\theta})$ is the probability to observe an event $x_{i}$, given the model $f$ and parameters $\bm{\theta}$. In the case where the PDF normalisation is a parameter of interest, the extended negative log likelihood is used
\begin{equation}
-\log\mathcal{L} = - \mu -N\log\mu + \log{N!} - \sum^{N}_{i=1}\log{f(x_{i} \mathbin{|} \bm{\theta})},
\label{eq:binnedlikelihood_ext}
\end{equation}
where $\mu$ is the expected number of events predicted by the model.
The additional terms have a negligible impact on the time of likelihood computations.\footnote{The constant terms are ignored by RooFit, as they are not needed to find the minimum, the other terms are only evaluated once.}
From \cref{eq:binnedlikelihood,eq:binnedlikelihood_ext}, it is evident that the time to compute a log likelihood scales with the number of events in the data set.
\begin{equation}
    T\left[ \mathrm{unbinned} \right] = 
    N_\mathrm{event} \cdot 
    T[\log{f(x_{i} \mathbin{|} \bm{\theta})}] = 
    \mathcal{O}(N_\mathrm{event}).
\end{equation}

\noindent
Analogously, the likelihood to fit binned data
\begin{equation}
\mathcal{L} = \prod^{N_\mathrm{bin}}_{i=1}\frac{\mu_{i}^{n_{i}}(\bm{\theta})}{n_{i}!} \cdot \exp\left(\mu_{i}(\bm{\theta})\right)
\label{equation:binned:binnedlikelihood}
\end{equation}
scales as
\begin{equation}
    T\left[ \mathrm{binned} \right] = 
    N_\mathrm{bin} \cdot 
    T[\log{\mu_{i}(\bm{\theta})}] = 
    \mathcal{O}(N_\mathrm{bin}).
\end{equation}
where $n_{i}$ is the number of events observed in bin $i$, and $\mu_{i}(\bm{\theta})$ the expected number of events in that bin. Computing $\mu$  takes roughly the same time as computing one event likelihood in the unbinned case, because it requires evaluating the model in the bin: $f(x_\mathrm{centre} \mathbin{|} \bm\theta)$.

By default, \roofit minimises negative log likelihoods using the MIGRAD algorithm of the MINUIT package~\cite{minuit}, and covariance matrices are estimated using the HESSE algorithm.
Since both MIGRAD and HESSE rely on calculating finite differences, their time complexities scale with the complexity of the likelihood calculations. 
The speed up when using binned fits is therefore proportional to $N_\mathrm{event} / N_\mathrm{bin}$.
Despite the fact that $T[\log{\mu_{i}(\bm{\theta})}; \text{bin sampling}] \approx 20 \cdot T[\log{f(x_{i} \mathbin{|} \bm{\theta})}]$ with our proposed solution and default integrator settings, since 21 points have to be evaluated to integrate each bin, binned fits are usually invoked for large data sets where $N_\mathrm{bin} \ll N_\mathrm{event}$. Hence, the solution presented in this paper still offers orders of magnitude speed up compared to unbinned fits, while eliminating unwanted biases.

\section{Validation with realistic pseudo experiments}

In order to demonstrate the real-world impact of these biases and the performance of our method, we perform pseudo experiments using examples representative of LHCb~\cite{Alves:2008zz} and ATLAS~\cite{Aad:2008zzm} collaboration analyses.

\subsection{LHCb charm example}
\label{sec:lhcb_charm}

\begin{figure}[tbp]
    \centering
    \begin{subfigure}[b]{0.48\linewidth}
        \centering
        \includegraphics[width=\linewidth,
            trim={0} {5mm} {5mm} {5mm},
            clip=true ]{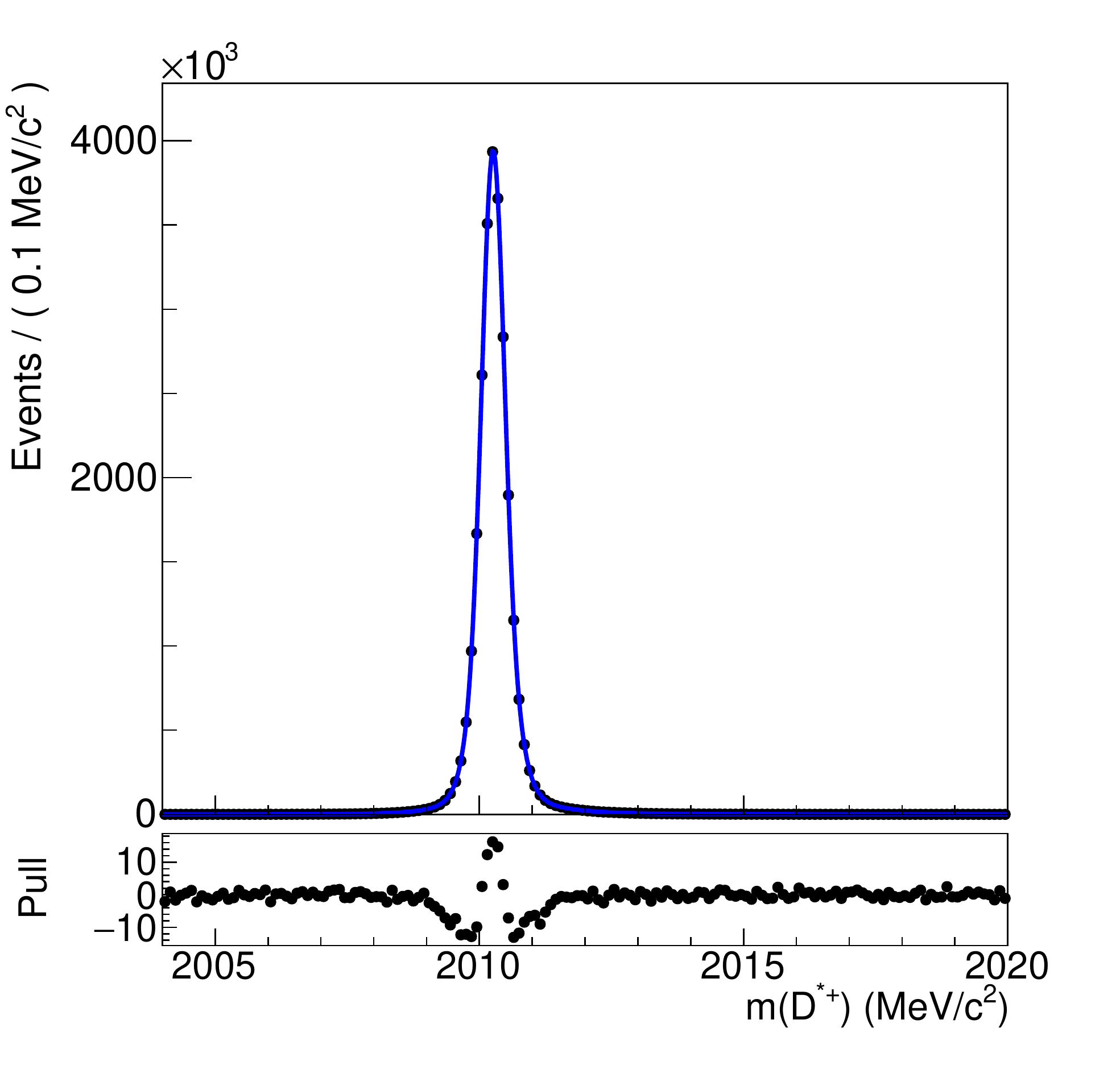}
        \caption{Default method}
    \end{subfigure}
    \hfil
    \begin{subfigure}[b]{0.48\linewidth}
        \centering
        \includegraphics[width=\linewidth,
            trim={0} {5mm} {5mm} {5mm},
            clip=true ]{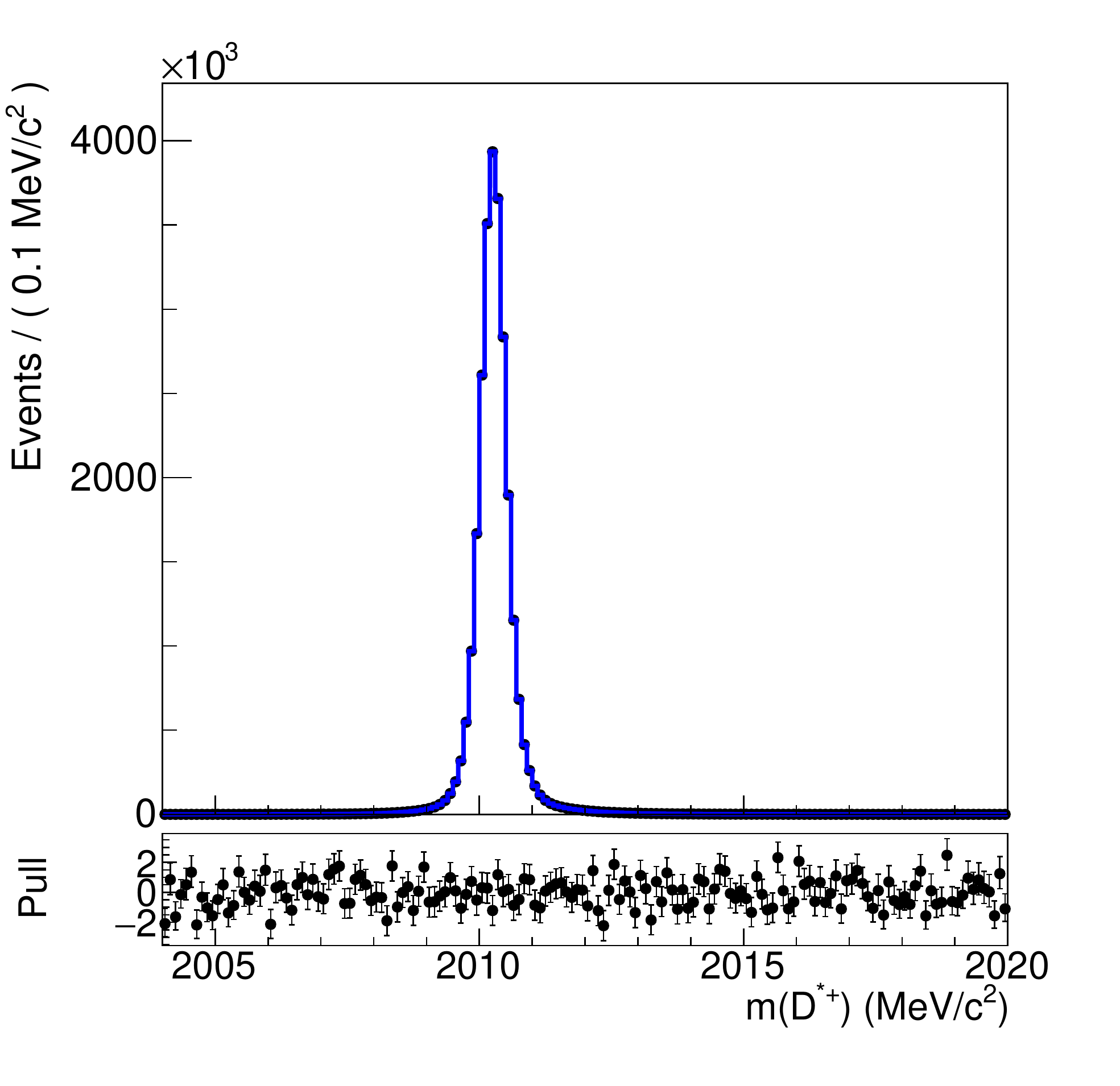}
        \caption{\textsc{RooBinSamplingPdf}}
    \end{subfigure}
    \caption{Example of a fit to one pseudo experiment for both methods. The bottom pads show pulls that are computed by comparing data counts with the plotted curves evaluated at the bin centres.}
    \label{fig:LHCb-charm-toyfits}
\end{figure}

We demonstrate and evaluate the performance of the default \roofit binned fit method and the \textsc{RooBinSamplingPdf} using the signal model and data set conditions (yield, number of bins) from~\cite{LHCb-PAPER-2020-045}. The PDF models the $D^{*+}$ peak of the decay $D^{*+} \to D^0(\to K^-\pi^+) \pi^+$ and is comprised of two Gaussian functions and a JohnsonSU distribution. The number of floating parameters is 10. While the total signal yield is \num{520E6} events, the fit is performed separately in 20 equipopulous bins of decay time, yielding \num{26E6} events to be fitted in one sample. The number of bins is 160. 

For each of the two methods, we generate $10^4$ pseudo experiments using the PDF values from one of the fits in reference~\cite{LHCb-PAPER-2020-045}. The generated data set is then fitted using the PDF with floating parameters. Example fits for each method are shown in \cref{fig:LHCb-charm-toyfits}. As in \cref{fig:comparisonBinSamplingPdf}, the bias is clearly seen when using the default fit method, while not so when using \textsc{RooBinSamplingPdf}. For each pseudo experiment and each fit parameter, we compute the difference between the fitted and generated value, divided by the error on the fitted value. While this quantity is not the exact equivalent of the statistical pull, as we use the generated value and not the expected one, the resulting distribution is still expected to follow a Gaussian with mean at 0 and standard deviation 1.

\begin{figure}[tbp]
    \centering
    \begin{subfigure}[b]{0.49\textwidth}
        \centering
        \includegraphics[width=\textwidth,%
          clip=true,%
          trim={0mm} {13mm} {0mm} {8mm} ]{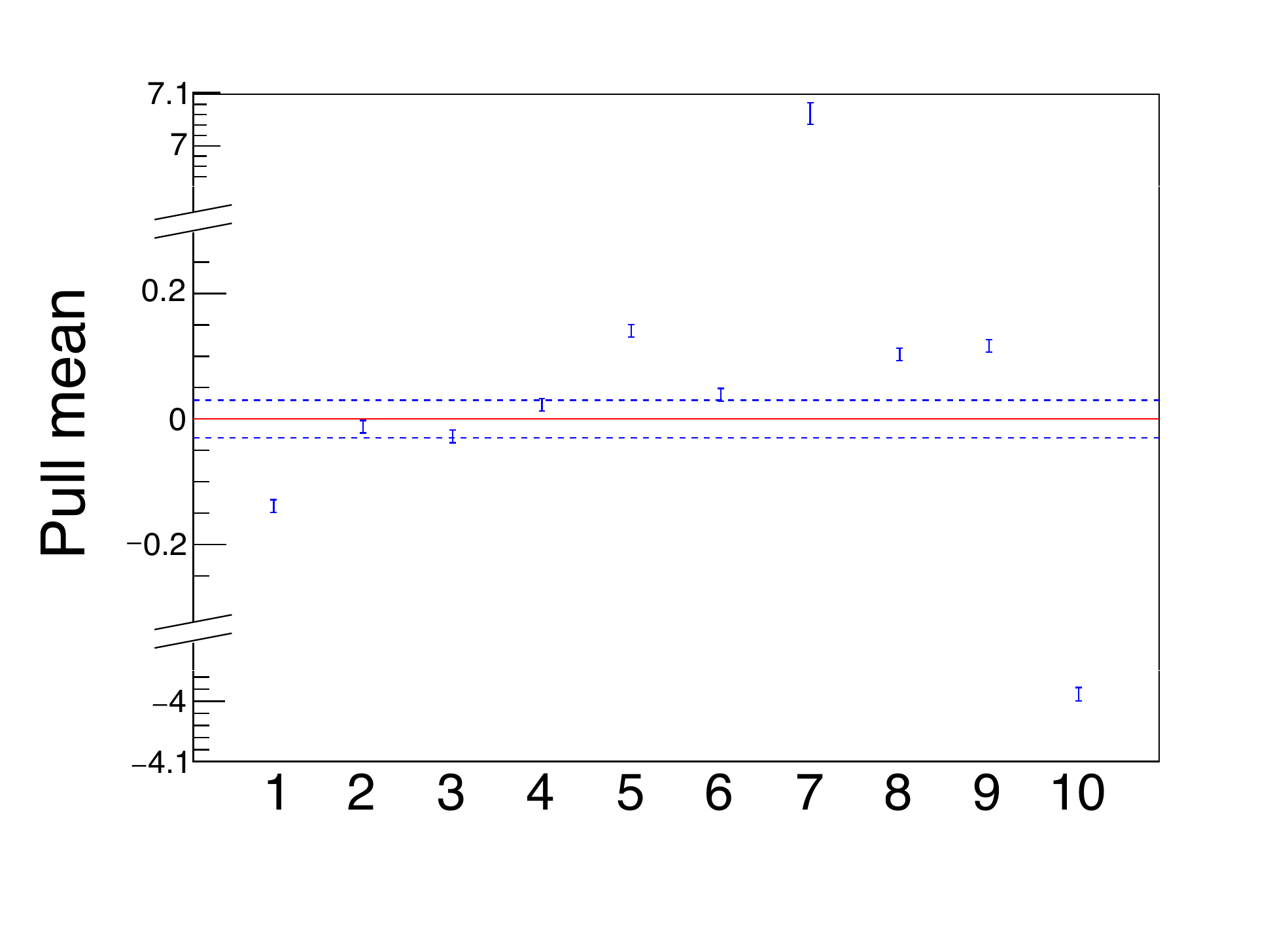}
        \caption{Default method: mean}
        \label{fig:pull_orig_mean}
    \end{subfigure}
    \hfil
    \begin{subfigure}[b]{0.49\textwidth}
        \centering
        \includegraphics[width=\textwidth,%
        clip=true,%
        trim={0mm} {13mm} {0mm} {8mm} ]{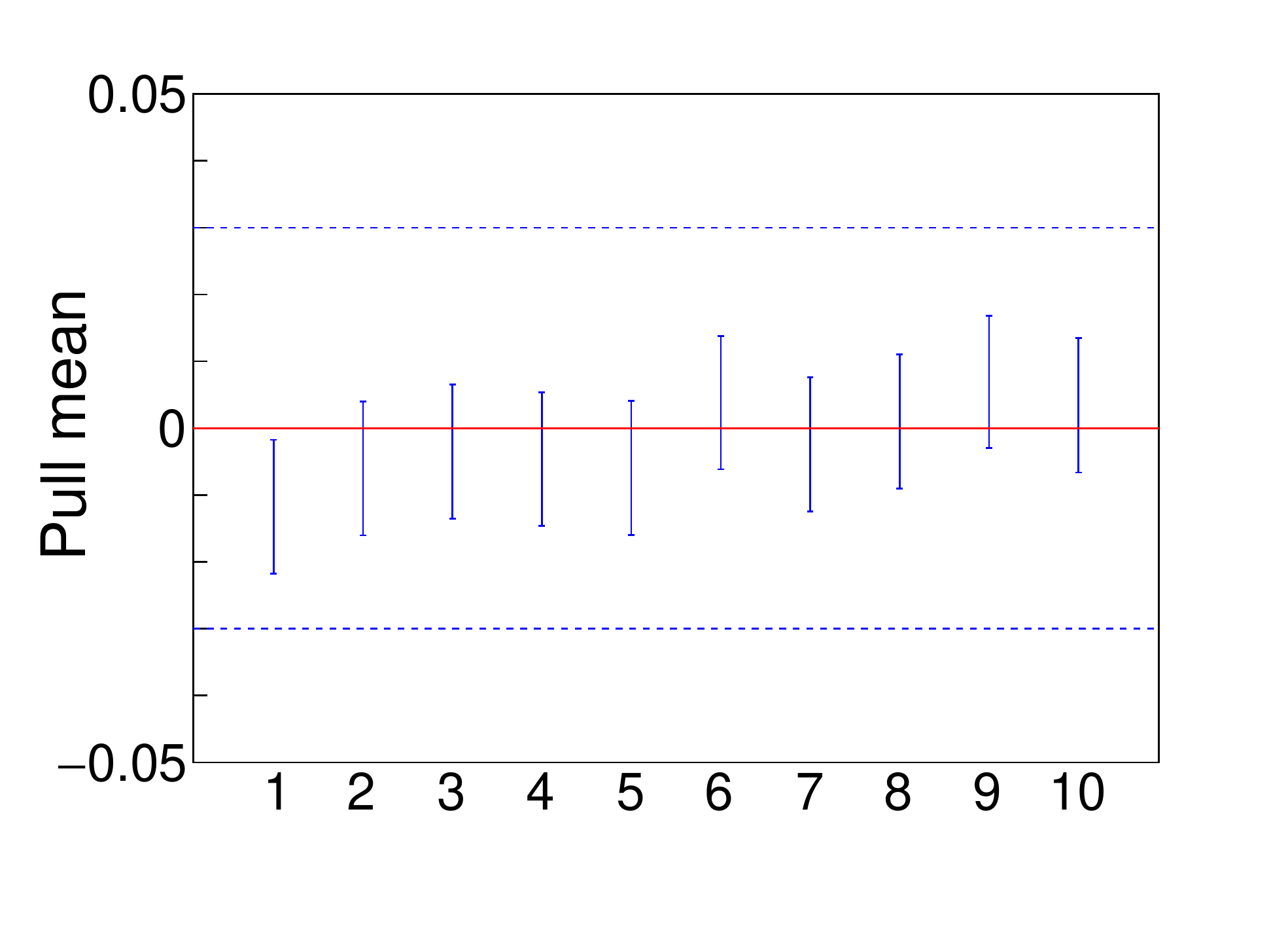}
        \caption{\textsc{RooBinSamplingPdf}: mean}
        \label{fig:pull_binSampling_mean}
    \end{subfigure}\\
    \begin{subfigure}[b]{0.49\textwidth}
        \centering
        \includegraphics[width=\textwidth,%
          clip=true,%
          trim={0mm} {13mm} {0mm} {8mm} ]{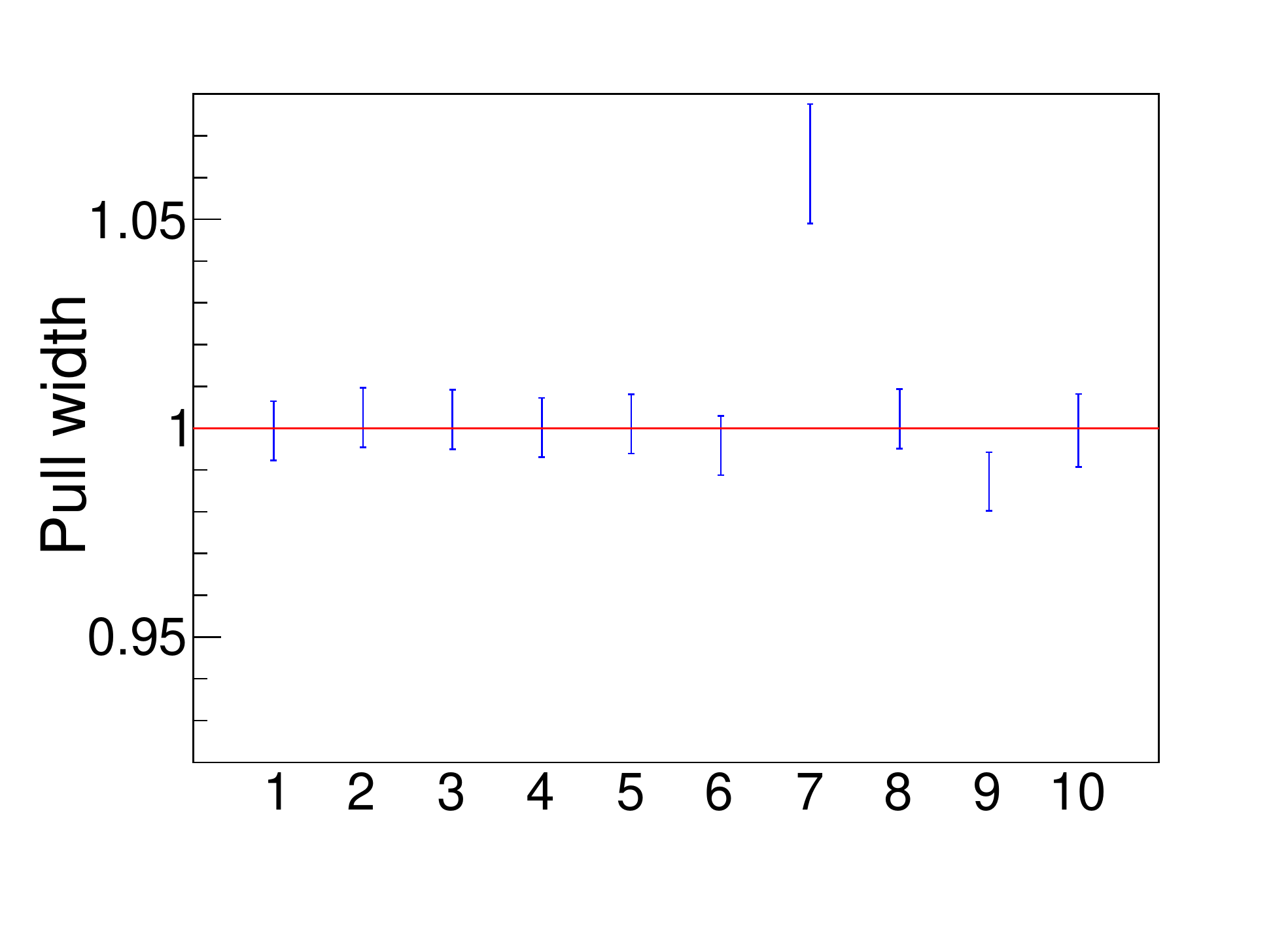}
        \caption{Default method: width}
    \end{subfigure}
    \hfil
    \begin{subfigure}[b]{0.49\textwidth}
        \centering
        \includegraphics[width=\textwidth,%
          clip=true,%
          trim={0mm} {13mm} {0mm} {8mm} ]{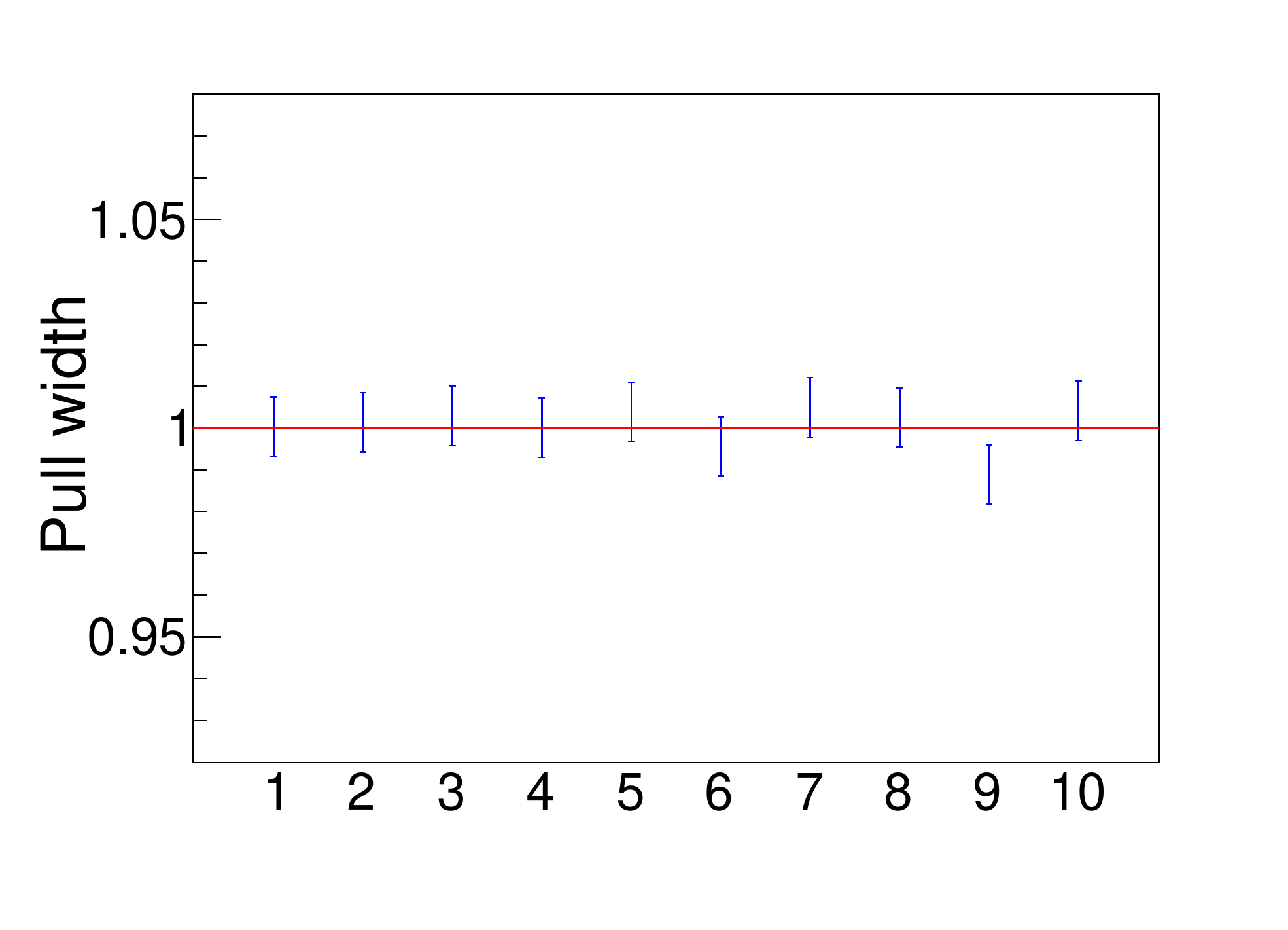}
        \caption{\textsc{RooBinSamplingPdf}: width}
    \end{subfigure}
    \caption{Mean (top) and width (bottom) of the pull of various fit parameters ($x$-axis), with respect to the generated and fitted values, for both methods. The blue dashed lines in (\subref{fig:pull_orig_mean}) and (\subref{fig:pull_binSampling_mean}) are at the same values to serve as visualisation aid.}
    \label{fig:LHCb-charm-pulls}
\end{figure}

We thus fit the obtained distribution with a Gaussian function and study the mean and standard deviation. The results are shown in \cref{fig:LHCb-charm-pulls}. It is clear that when using the default method, large biases are present in almost all fitted parameters. Contrary, for the \textsc{RooBinSamplingPdf}, the fitted parameters show no discernible bias. The plots of the standard deviation show that the error estimation is reliable for all parameters when using \textsc{RooBinSamplingPdf}, and all but one for the default method.

\medskip

\begin{figure}[t]
    \centering
    \includegraphics[width=0.75\textwidth,%
      clip=true,%
      trim={0mm} {0mm} {10mm} {10mm} ]{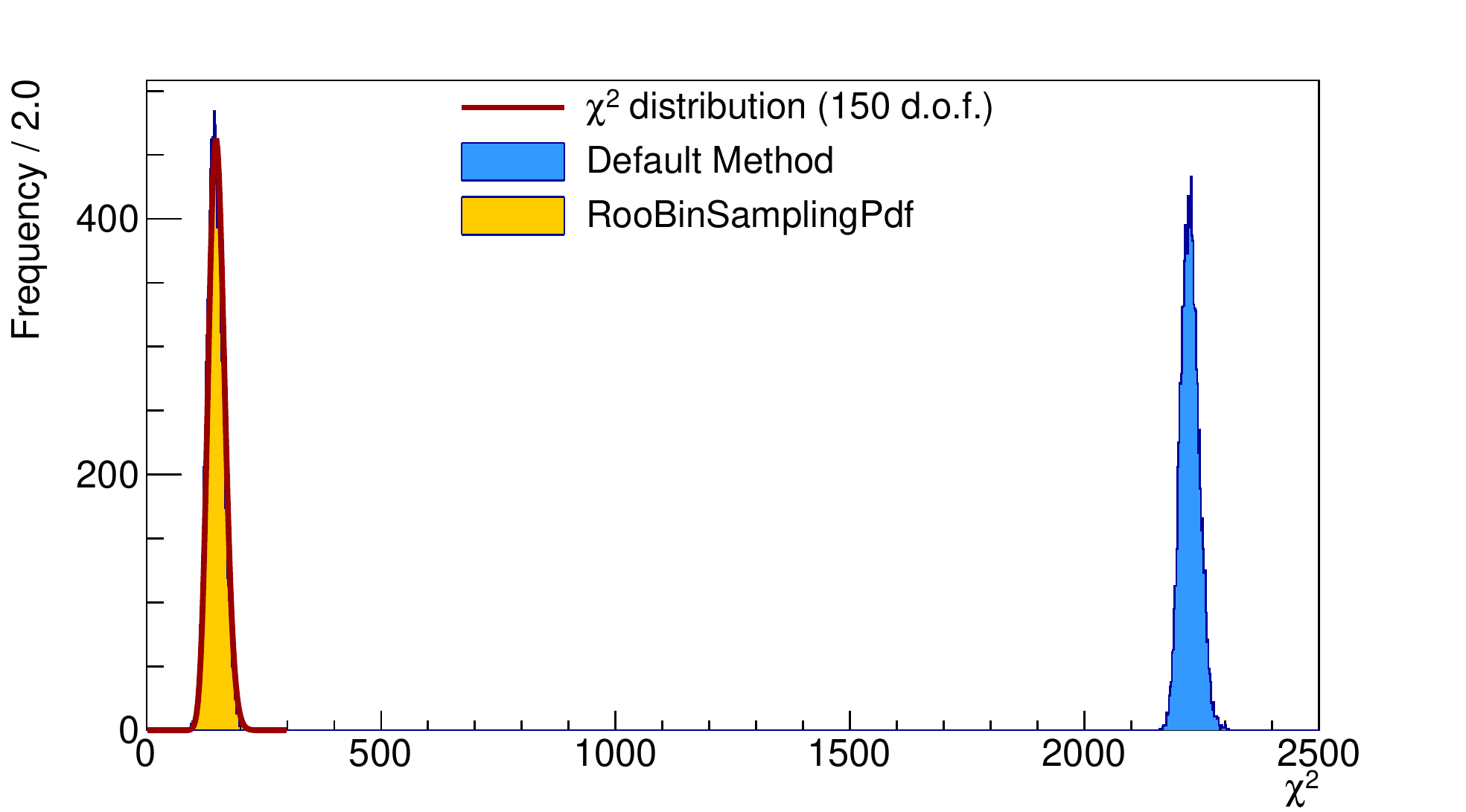}
    \caption{Comparison of the theoretical $\chi^2$ distribution with NDOF = 150 and the empirical distributions obtained from fits to pseudo experiments for both methods. While the standard method results in very
    high values of $\chi^2$, the values with \textsc{RooBinSamplingPdf} follow the theoretical distribution.}
    \label{fig:LHCb-charm-chi2}
\end{figure}

As an additional cross-check, we examine the distribution of $\chi^2$ values obtained from the fits of the pseudo experiments. The number of degrees of freedom is the number of bins minus the number of free parameters in the fit, i.e.\ 150, so the obtained distribution is compared to the theoretical $\chi^2$ distribution for 150 degrees of freedom. The results are shown in \cref{fig:LHCb-charm-chi2}. We can observe that when using \textsc{RooBinSamplingPdf}, the obtained distribution matches reasonably well the theoretical one, while for the default binned fit method, it is very different. 

\begin{figure}[tb]
    \centering
    \begin{subfigure}[b]{0.49\textwidth}
        \centering
        \includegraphics[width=\textwidth,%
         clip=true,%
         trim={0mm} {13mm} {0mm} {8mm}
        ]{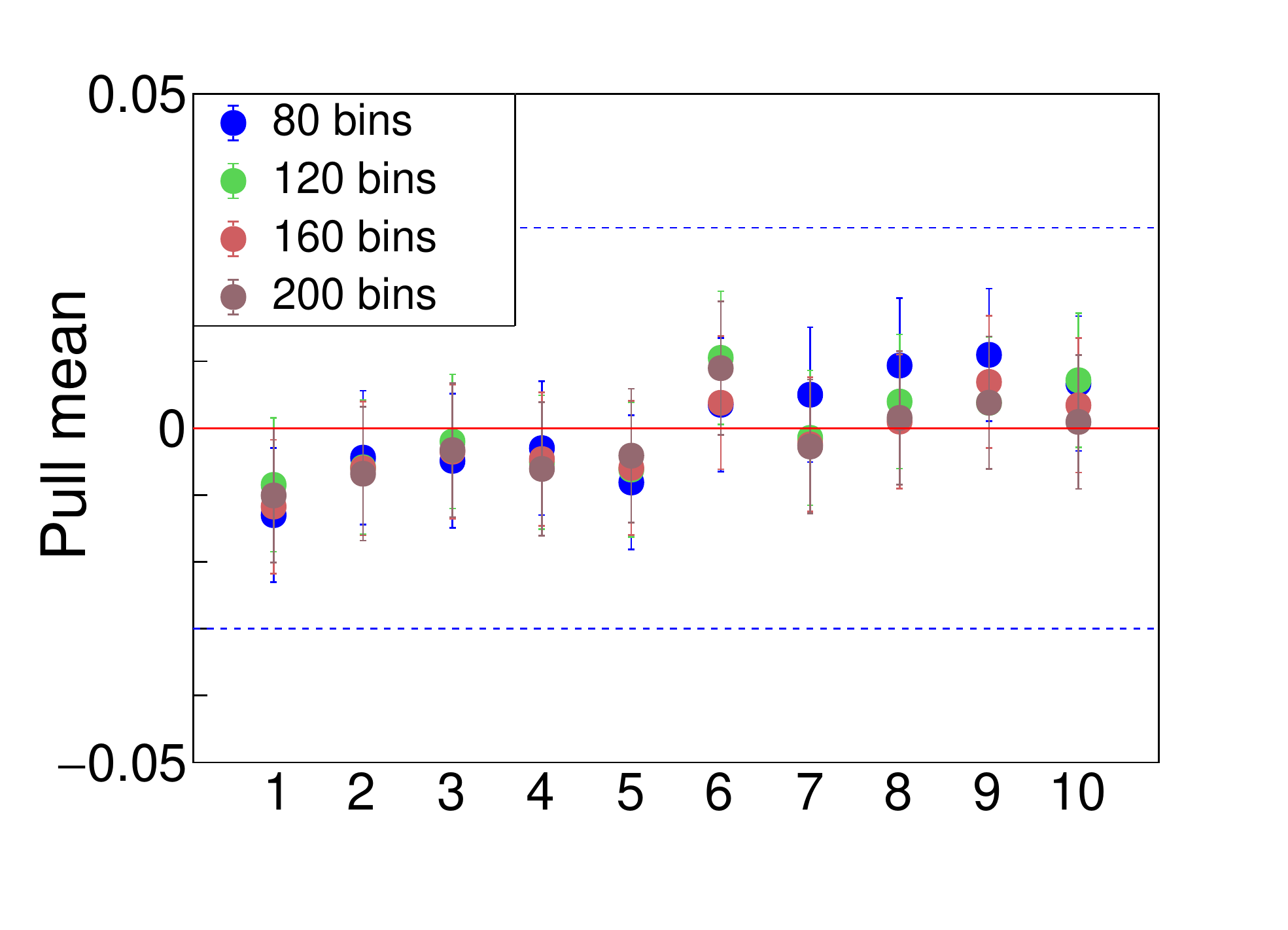}
        \caption{Mean}
    \end{subfigure}
    \begin{subfigure}[b]{0.49\textwidth}
        \centering
        \includegraphics[width=\textwidth,%
         clip=true,%
         trim={0mm} {13mm} {0mm} {8mm} ]{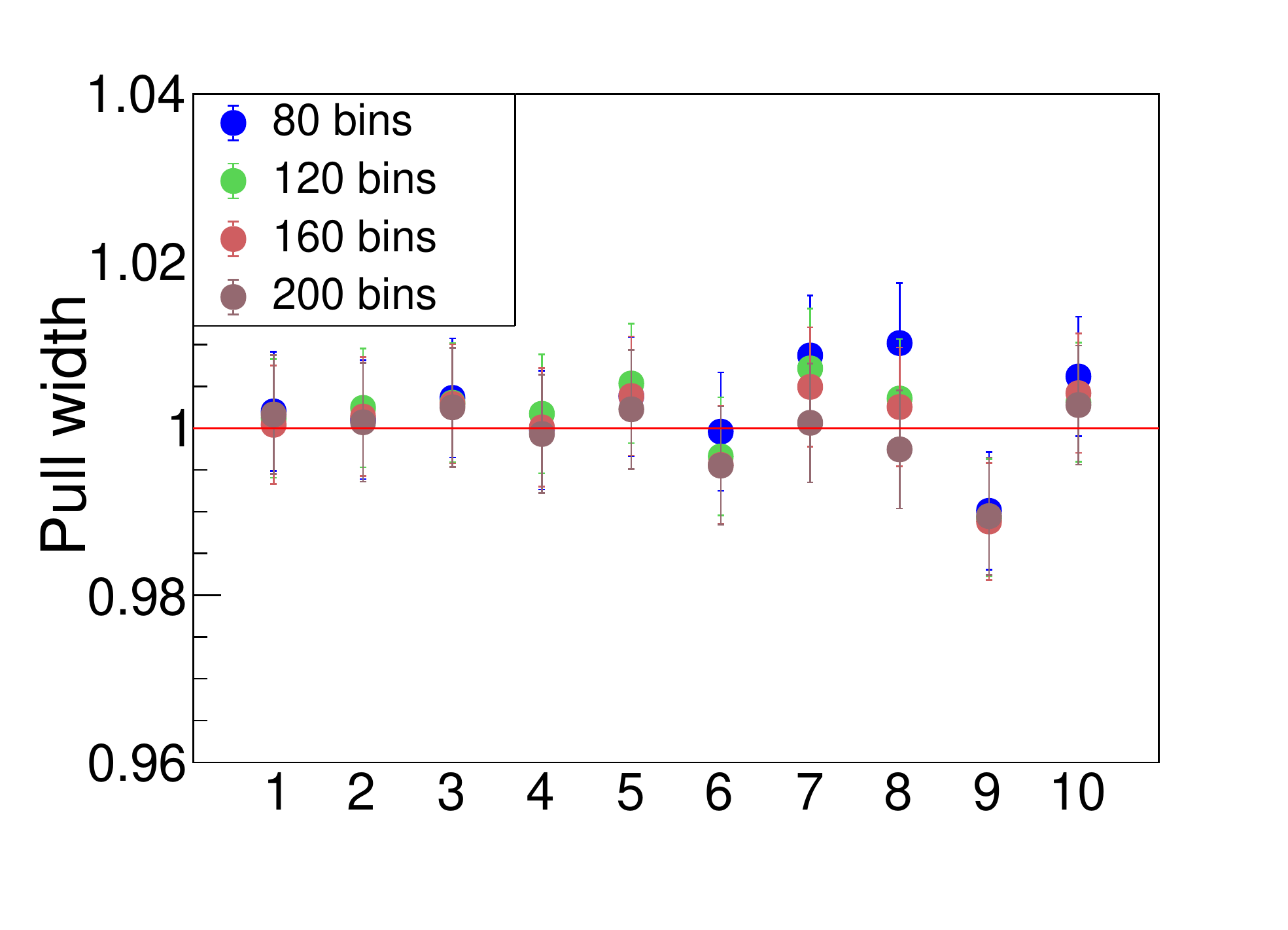}
        \caption{Width}
    \end{subfigure}
    \caption{Mean and width of the pulls of various fit parameters ($x$-axis), with respect to the generated and fitted values, for different choices of binning using \textsc{RooBinSamplingPdf}.}
    \label{fig:LHCb-charm-bins}
\end{figure}

In the above described studies, the number of bins was taken as that reported in reference~\cite{LHCb-PAPER-2020-045}. \Cref{fig:LHCb-charm-bins} shows plots  of the mean and width of the Gaussian distribution, fitted to the ratio of the difference between the generated and fitted values and the error of the fitted value (analogue to \cref{fig:LHCb-charm-pulls}), for different binnings using \textsc{RooBinSamplingPdf}. It can be concluded that a different choice of binning does not affect the performance of the \textsc{RooBinSamplingPdf}.

\pagebreak

\subsection{LHCb beauty example}
\label{sec:lhcb_beauty}

\begin{figure}[t]
\centering
\includegraphics[width=0.5\linewidth,
    trim={0} {0} {5mm} {4mm}
    clip=true ]{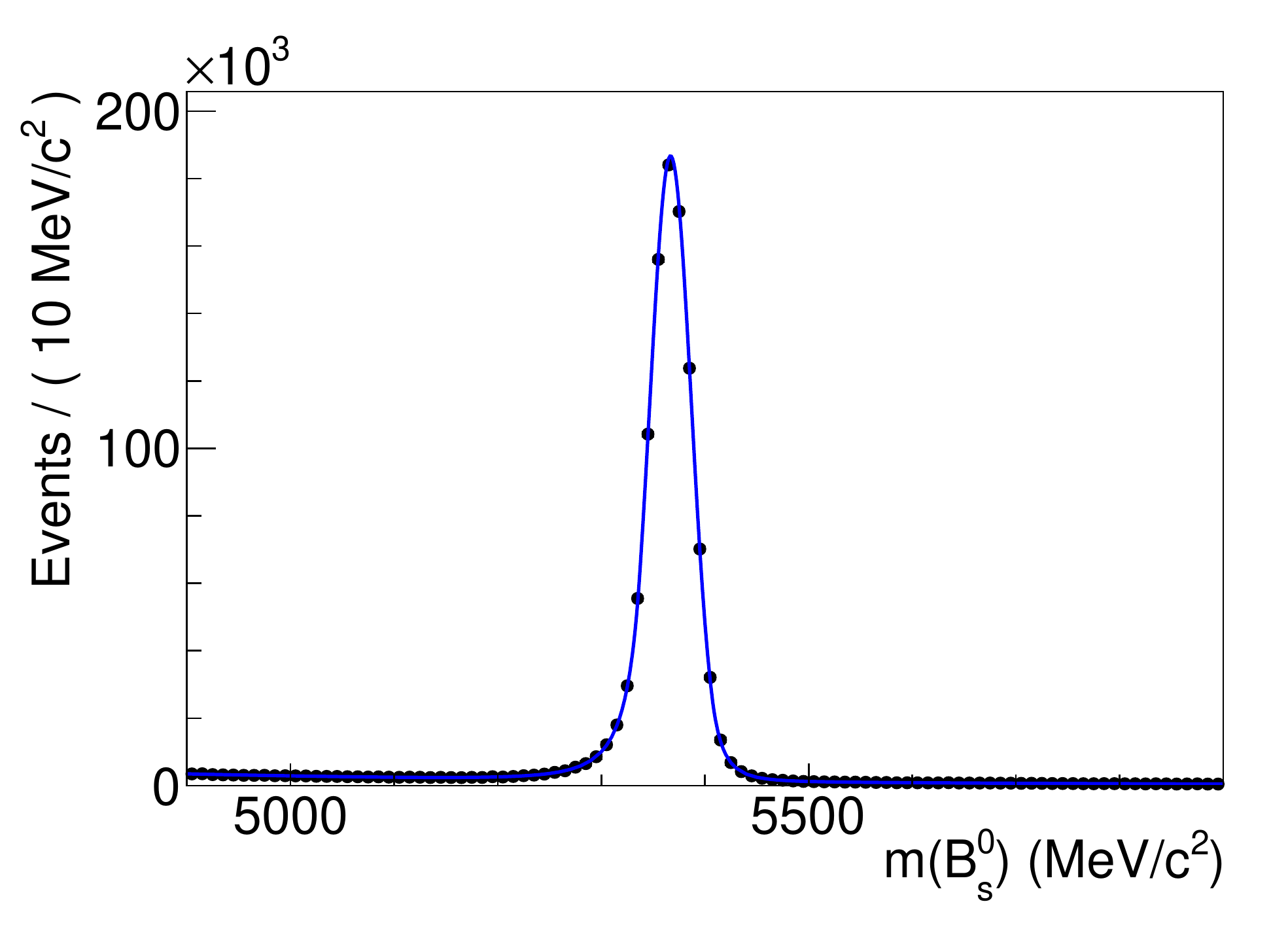}
\caption{An example pseudo experiment generated by the baseline mode. The data set (black) is generated by a model (blue) which consists of a DSCB signal and a combinatorial background.}
\label{figure:binned:generator}
\end{figure}

In this section, we explore in more detail the effects of the bias with regard to signal width, number of signal events and number of bins, using a model with a signal PDF similar to typical beauty meson decays at LHCb and a combinatorial background model.

The signal is a Double Sided Crystal Ball (DSCB) function \cite{Skwarnicki:1986xj}, with a Gaussian core of mean $\mu_{\mathrm{gauss}} = \SI{5366.79}{\mev}$ (corresponding to the \Bs rest mass) and width $\sigma_{\mathrm{gauss}} = \SI{20}{\mev}$. 
The combinatorial background is an exponential function with a slope of $b_{\mathrm{slope}} = \SI{-2E-3}{\mev^{-1}}$. The baseline model has \num{1E6} signal events and \num{1.5E5} background events within a mass window of \SIrange{4900}{5900}{\mev}. A plot of the typical data set generated by this model is shown in \cref{figure:binned:generator}. 


We study the bias as a function of signal width, the number of signal events, and the number of bins. At a given bin width, broader signal distributions are expected to be less biased than narrower ones. Similarly, fits to the same signal distribution with narrower bin widths are expected to be less biased than fits with wider bins.

\begin{figure}[tbhp]
\centering
\begin{minipage}[b]{0.49\linewidth}
\centering
\includegraphics[width=\linewidth]{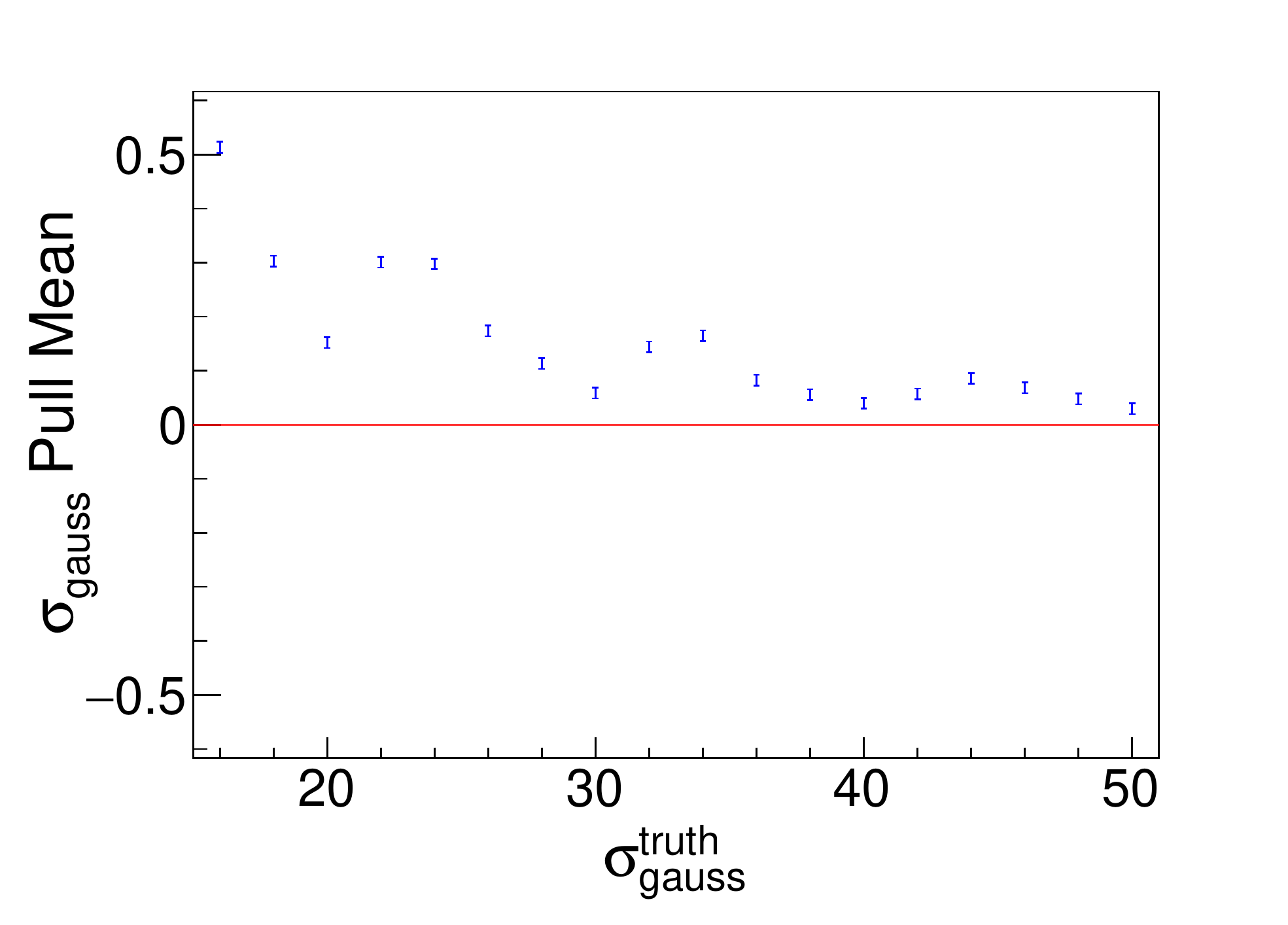}
\end{minipage}
\begin{minipage}[b]{0.49\linewidth}
\centering
\includegraphics[width=\linewidth]{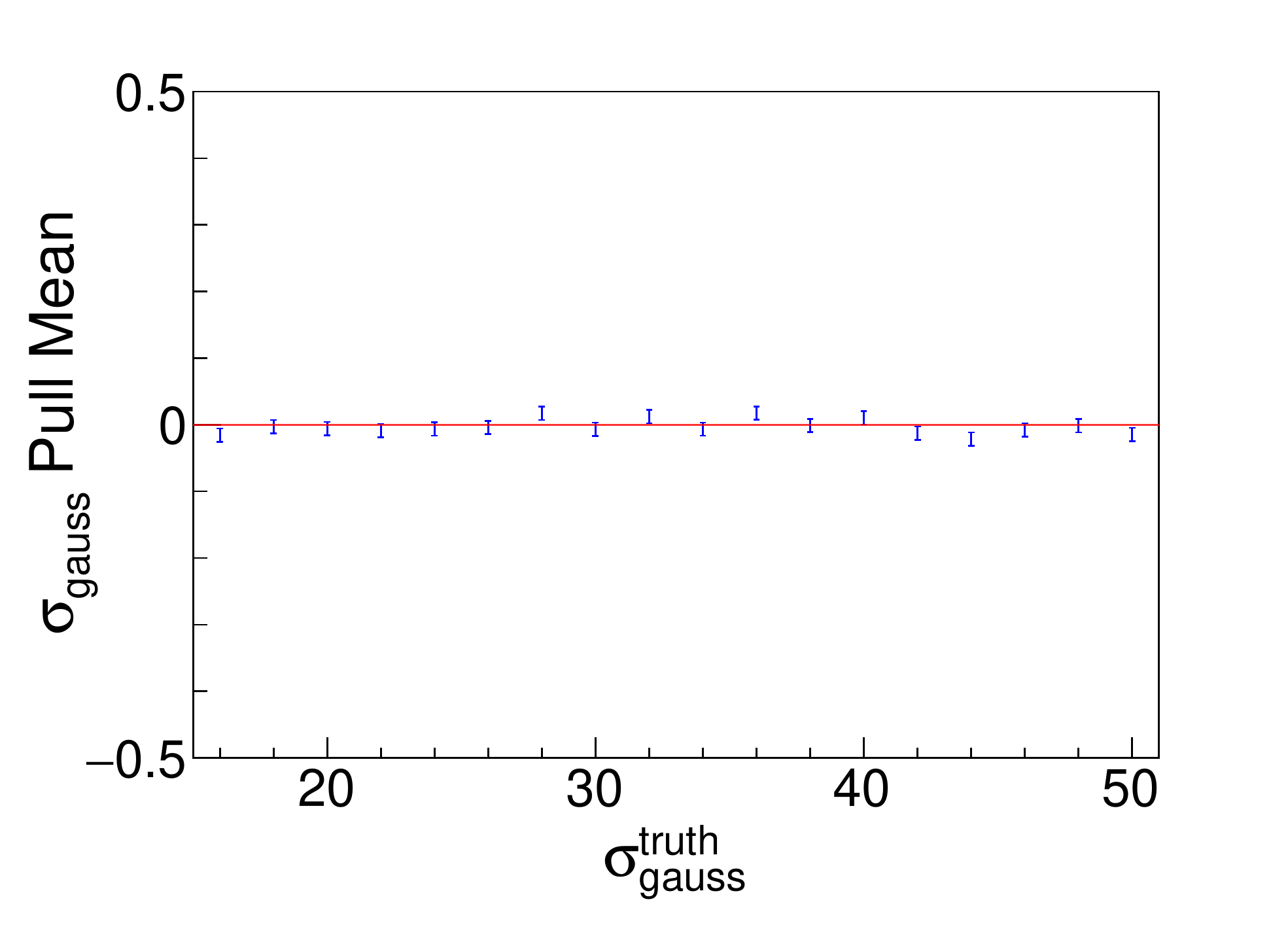}
\end{minipage}
\begin{minipage}[b]{0.49\linewidth}
\centering
\includegraphics[width=\linewidth]{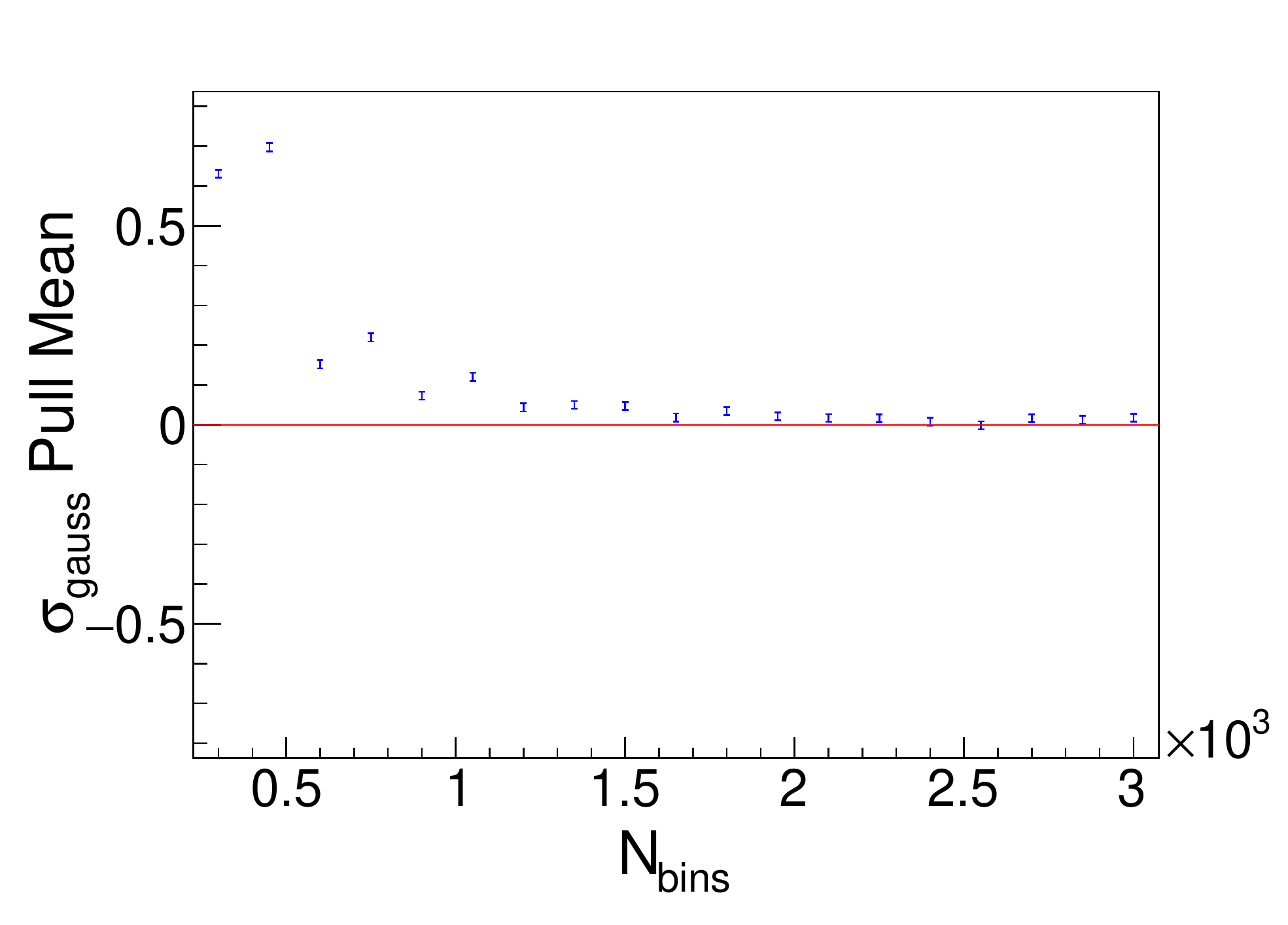}
\end{minipage}
\begin{minipage}[b]{0.49\linewidth}
\centering
\includegraphics[width=\linewidth]{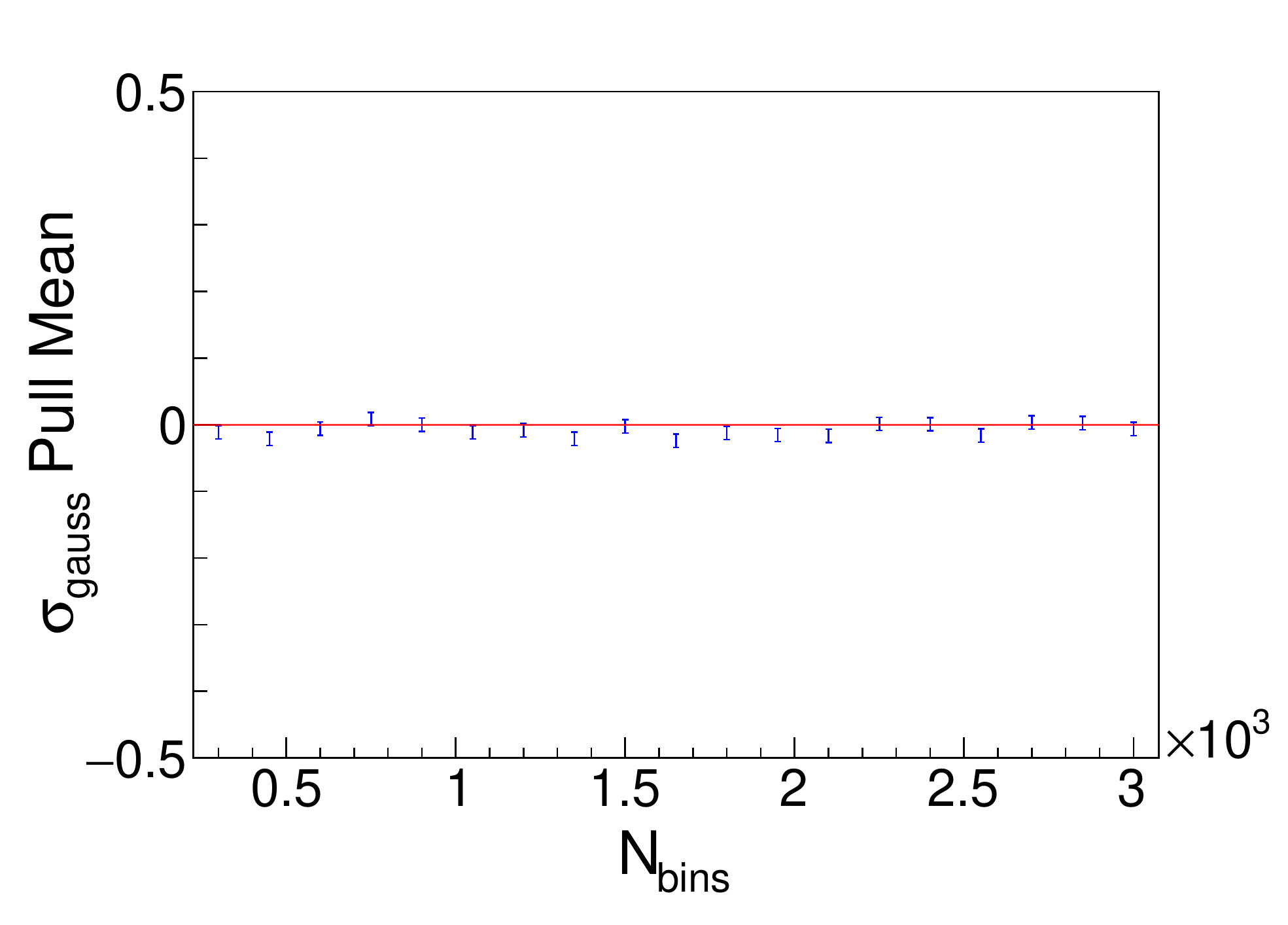}
\end{minipage}
\caption{Mean pull of $\sigma_\mathrm{gauss}$ in dependence of $\sigma_\mathrm{gauss}^\mathrm{truth}$ (top) and $N_\mathrm{bins}$ (bottom) with the default method (left) and \textsc{RooBinSamplingPdf} (right).}
\label{figure:binned:binnedResults}
\end{figure}

\begin{figure}[t]
\centering
\includegraphics[width=0.55\linewidth,
    trim={2mm} {3mm} {3mm} {12mm},
    clip=true ]{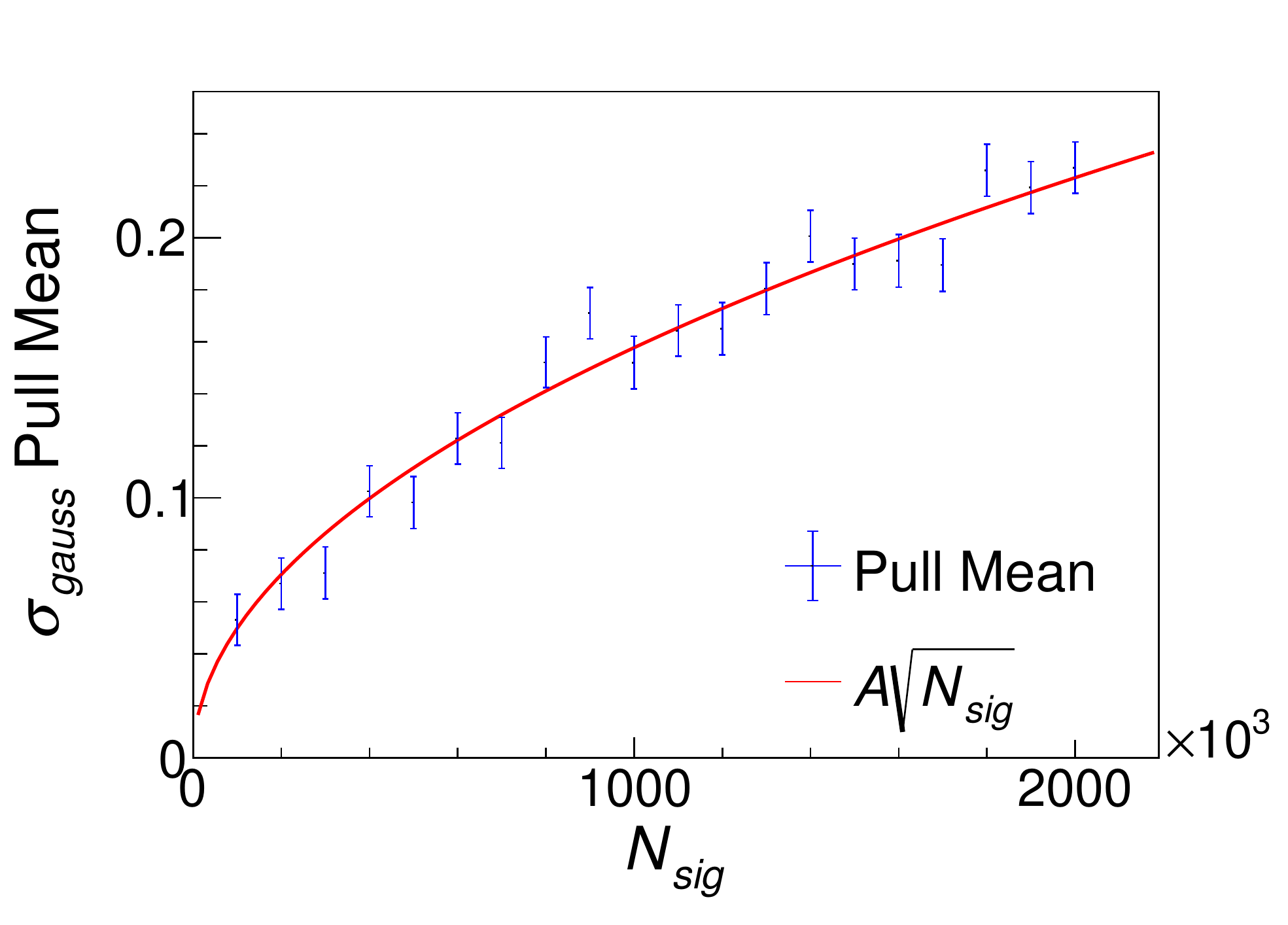}
\caption{The bias of $\sigma_\mathrm{gauss}$  against $N^\mathrm{truth}(\mathrm{signal})$ (black) is well modelled by $A\sqrt{N(\mathrm{signal})}$ (red).}
\label{figure:binned:binnednSig}
\end{figure}

The results are shown in \cref{figure:binned:binnedResults}. As we can see, the pull bias in $\sigma_\mathrm{gauss}$ increases as $\sigma_\mathrm{gauss}^\mathrm{truth}$ decreases and as $N_\mathrm{bins}$ decreases.  The $\sigma_\mathrm{gauss}$ bias is plotted as a function of $N^\mathrm{truth}(\mathrm{signal})$ in \cref{figure:binned:binnednSig}. Interestingly, the bias is well modelled as $A\sqrt{N(\mathrm{signal})}$. This proves that the absolute bias remains unchanged with respect to $N^\mathrm{truth}(\mathrm{signal})$, but it is the $\frac{1}{\sqrt{N(\mathrm{signal})}}$ uncertainty scaling that amplifies the pull bias in $\sigma_\mathrm{gauss}$.


\subsection{ATLAS example}

\begin{figure}[t]
\centering
\includegraphics[width=0.58\linewidth,
    trim={0} {0} {0} {8mm},
    clip=true ]{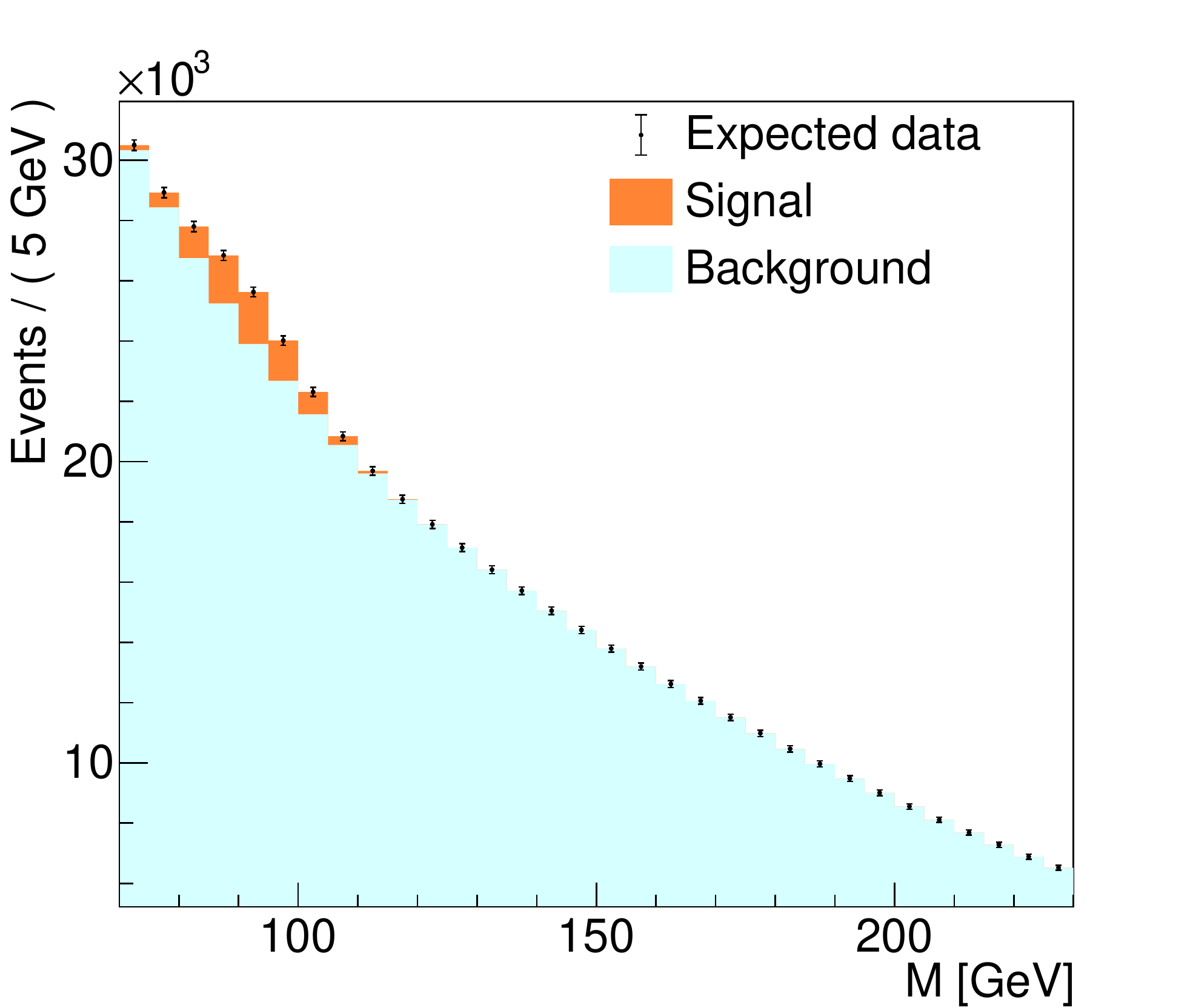}
\caption{The default signal-plus-background expected distribution fitted with the \textsc{RooBinSamplingPdf} integration. Expected data points are shown as black points, whereas the resulting fitted model consists of the signal (orange) stacked on top of the background (cyan).}
\label{fig:ATLAS-plot}
\end{figure}

The adoption of \textsc{RooBinSamplingPdf} functionalities finds extensive application in analyses carried out by the ATLAS Collaboration. In fact, some ATLAS analyses rely on analytic models and binned likelihood fits, see for example $H\to\mu\mu$~\cite{ATLASHmumu}. 
The absence of \textsc{RooBinSamplingPdf} forced many physics analyses to rely on very small bin widths, as the bias shown in \cref{fig:pulls_continuousPdf} becomes negligible with bin size orders of magnitude smaller than the resolution in the characteristic structures of the considered fit observable. 

However, analyses based on a mixture of pure analytic models and template PDFs from simulation may not be able to accommodate a full parameterisation of the fit model.
Recent ATLAS studies of the inclusive boosted $H\to b\bar{b}$ production at high Higgs-boson transverse momentum~\cite{ATLAS-CONF-2021-010} exploit the new class, object of this paper. The preliminary version of the work discussed in Ref.~\cite{ATLAS-CONF-2018-052} was completed before the presented \roofit developments and is, therefore, taken as a natural field of application for the \textsc{RooBinSamplingPdf} class. This example supplements studies shown in \cref{sec:lhcb_charm,sec:lhcb_beauty}, as the signal-to-background ratio is much smaller and an analytic model is exploited to describe the background only, inducing an indirect bias in the signal extraction.

The model consists of a linear sum of the ``background$"$ and ``signal$"$ models. The background is described by means of a polynomial exponential PDF model of the form:
\begin{equation*}
    f_{N}\left(x \,\middle|\,\vec{\theta}\,\right) = \displaystyle\exp\left(\sum_{i=1}^{N} \theta_{i} x^{i}\right),
\end{equation*}where $N=4$, $x=(M-150)/80$, with $M$ the fit observable, and $\theta_i$ are fit parameters. A fit range of [70,230] is taken for $M$, resulting in [-1,1] for $x$. The following model parameter values are taken: $\vec{\theta}=(-0.7,-0.05,-0.1,0.05)$.
The signal shape is fixed and modelled via a Gaussian centred at $M=91$ with a standard deviation of $\sigma=8.5$.
The default signal fraction (\fsig), total number of events (\Nev) and of bins (\Nbins) are \numlist{0.015; 5E5; 32}, respectively.
\Cref{fig:ATLAS-plot} shows the resulting falling background spectrum with a ``$Z$-like$"$ resonance on top.

The bias in the extracted fit parameters is dependent on the global and local statistical accuracy and is, therefore, studied as a function of the available statistics over the whole fit range (\Nev) and in each fit bin (\Nbins). For each value of \Nev and \Nbins a thousand data sets are generated through Poisson-sampling of the expected distribution, and fitted with both the default method and the \textsc{RooBinSamplingPdf}. 

\begin{figure}[hbtp]
    \centering
    \begin{subfigure}[b]{0.39\textwidth}
        \centering
        \includegraphics[width=\textwidth]{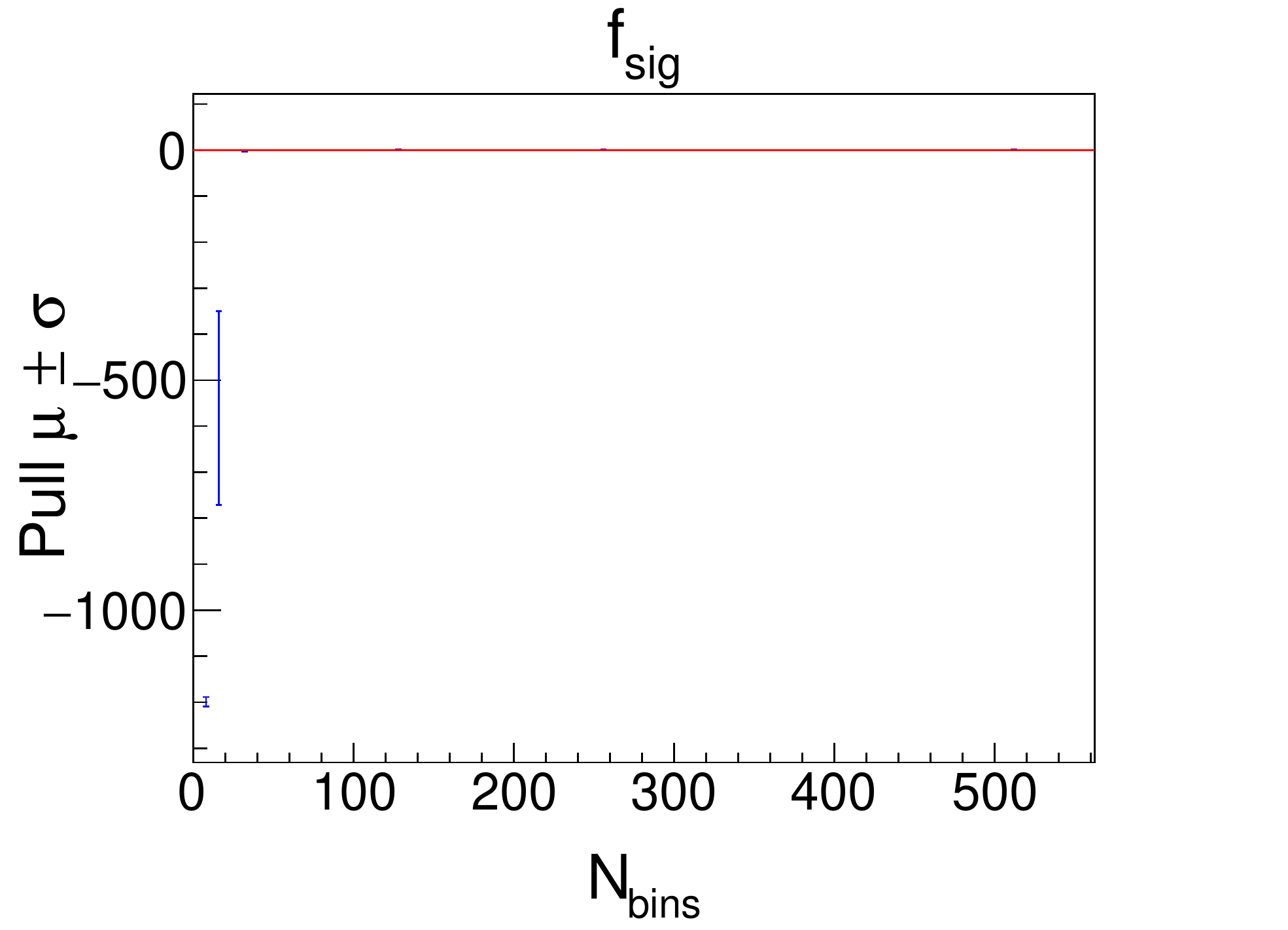}
    \end{subfigure}
    \begin{subfigure}[b]{0.39\textwidth}
        \centering
        \includegraphics[width=\textwidth]{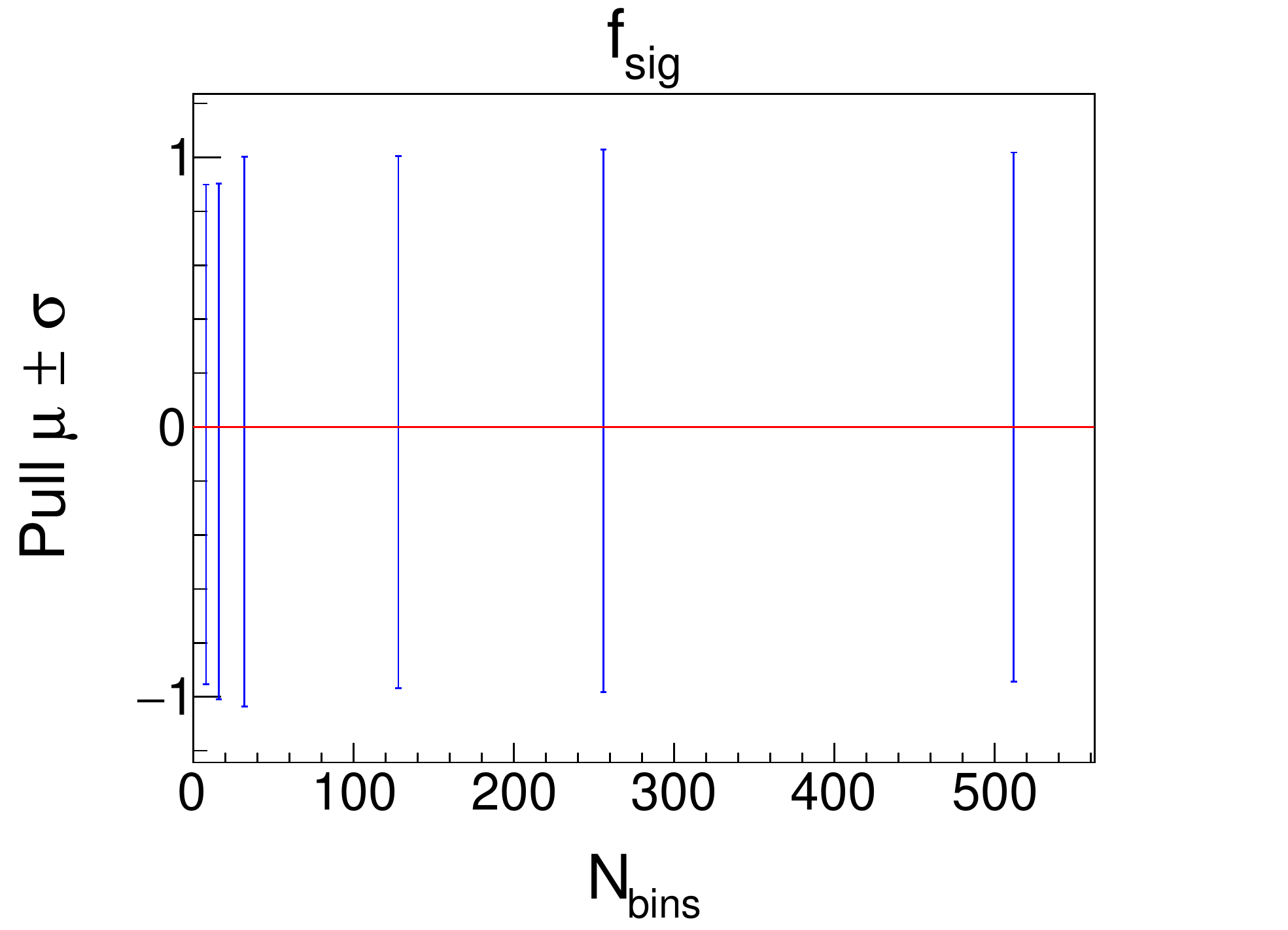}
    \end{subfigure}\\
    \begin{subfigure}[b]{0.39\textwidth}
        \centering
        \includegraphics[width=\textwidth]{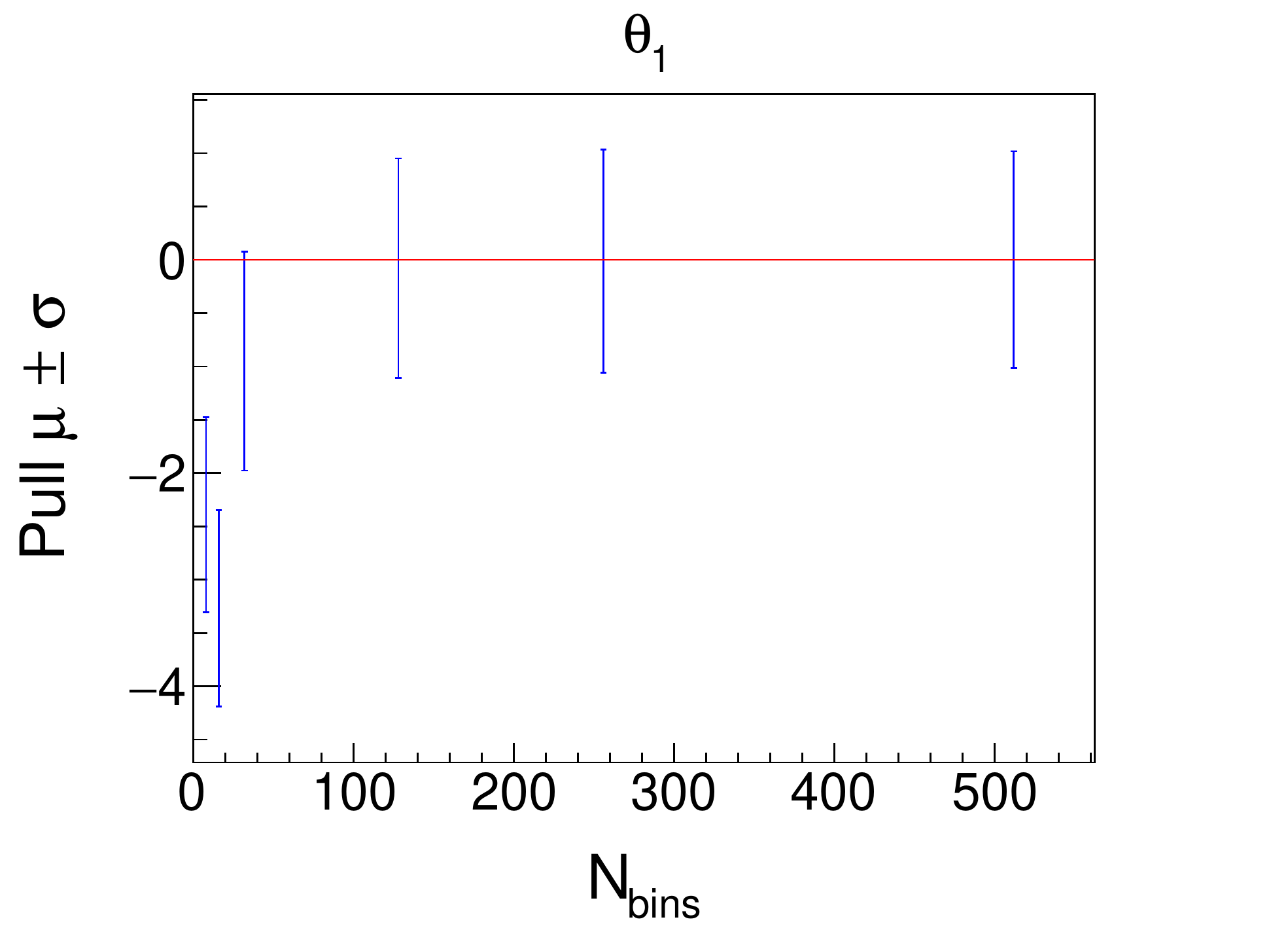}
    \end{subfigure}
    \begin{subfigure}[b]{0.39\textwidth}
        \centering
        \includegraphics[width=\textwidth]{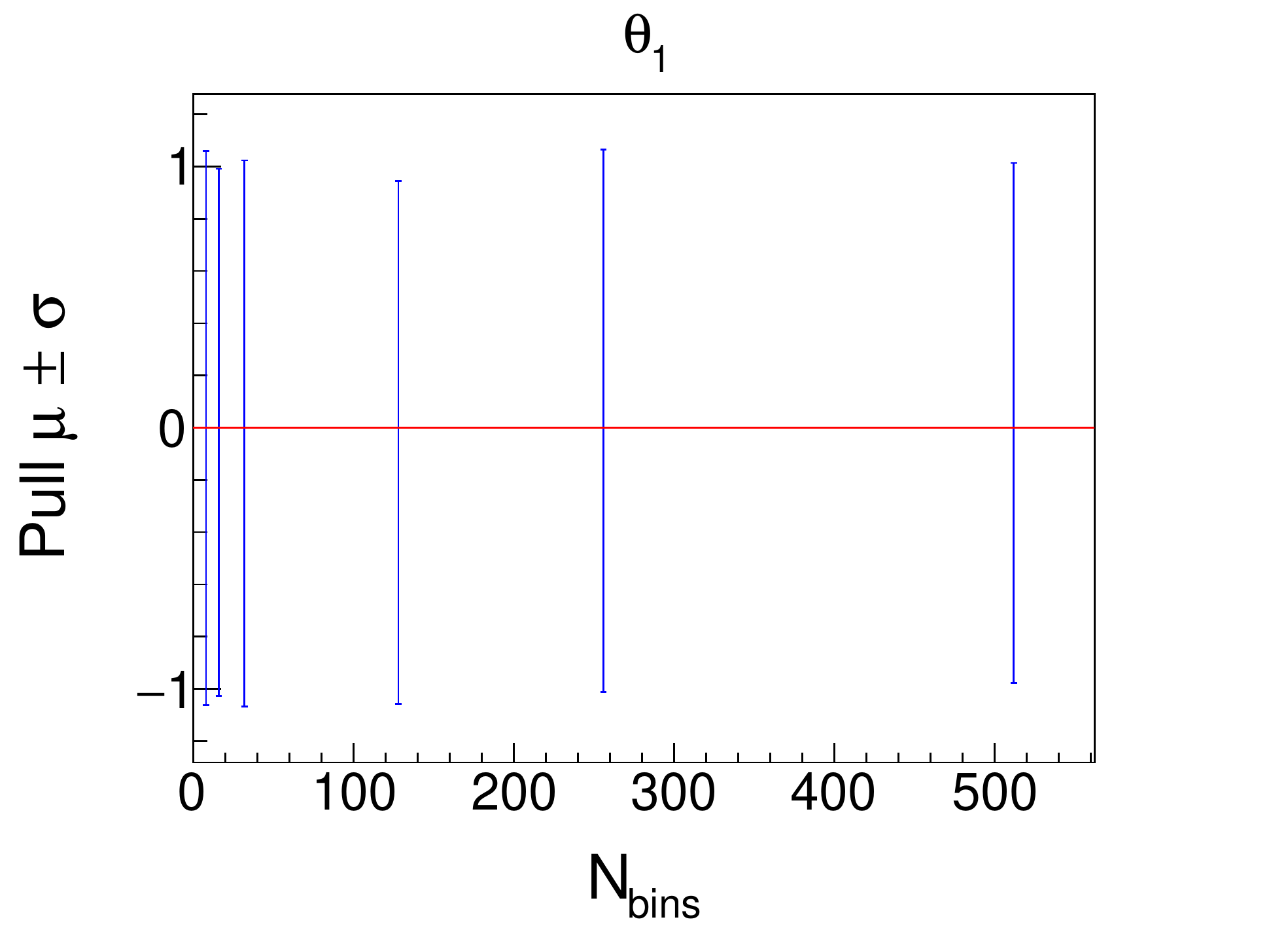}
    \end{subfigure}
    \begin{subfigure}[b]{0.39\textwidth}
        \centering
        \includegraphics[width=\textwidth]{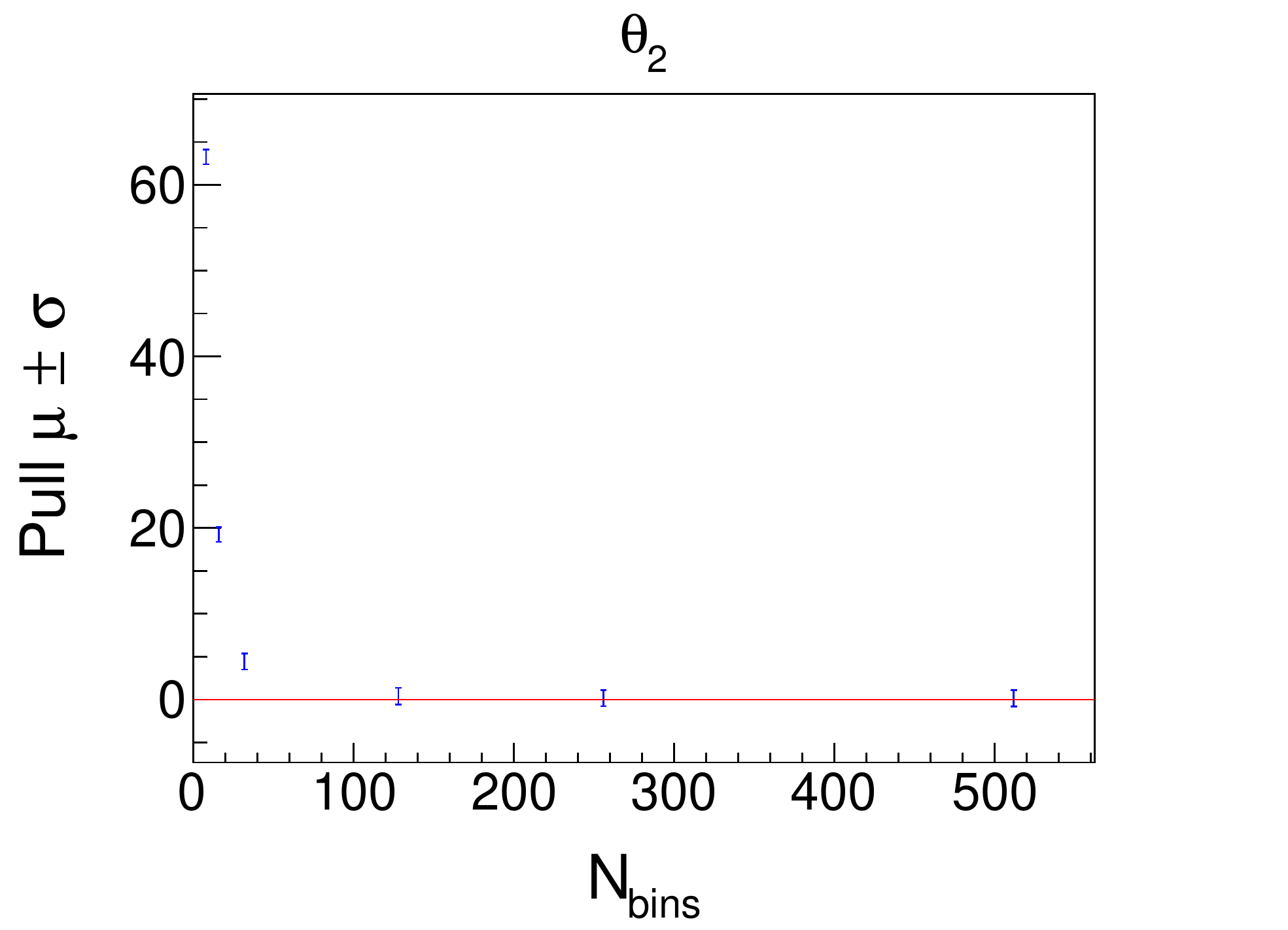}
    \end{subfigure}
    \begin{subfigure}[b]{0.39\textwidth}
        \centering
        \includegraphics[width=\textwidth]{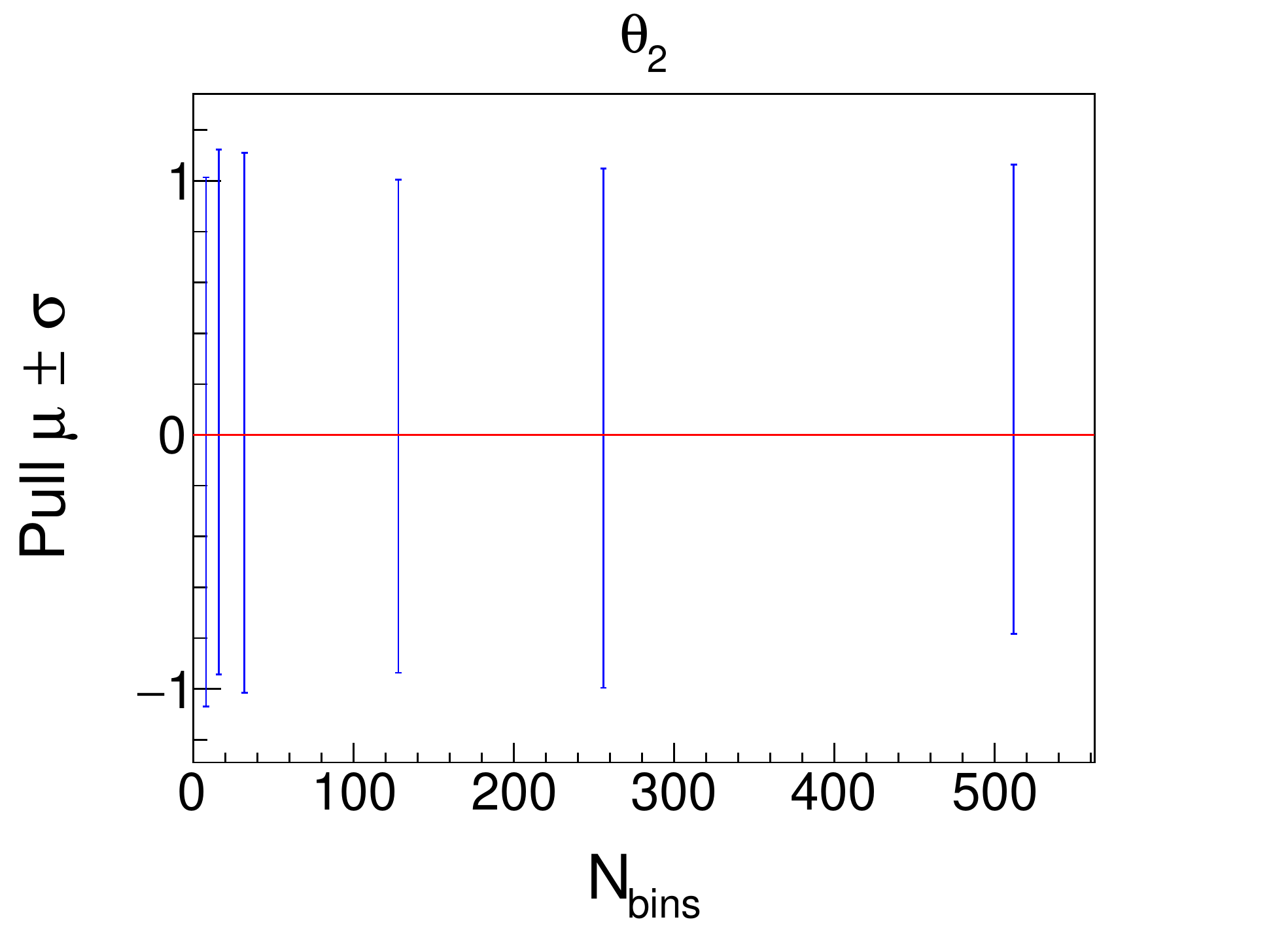}
    \end{subfigure}
    \begin{subfigure}[b]{0.39\textwidth}
        \centering
        \includegraphics[width=\textwidth]{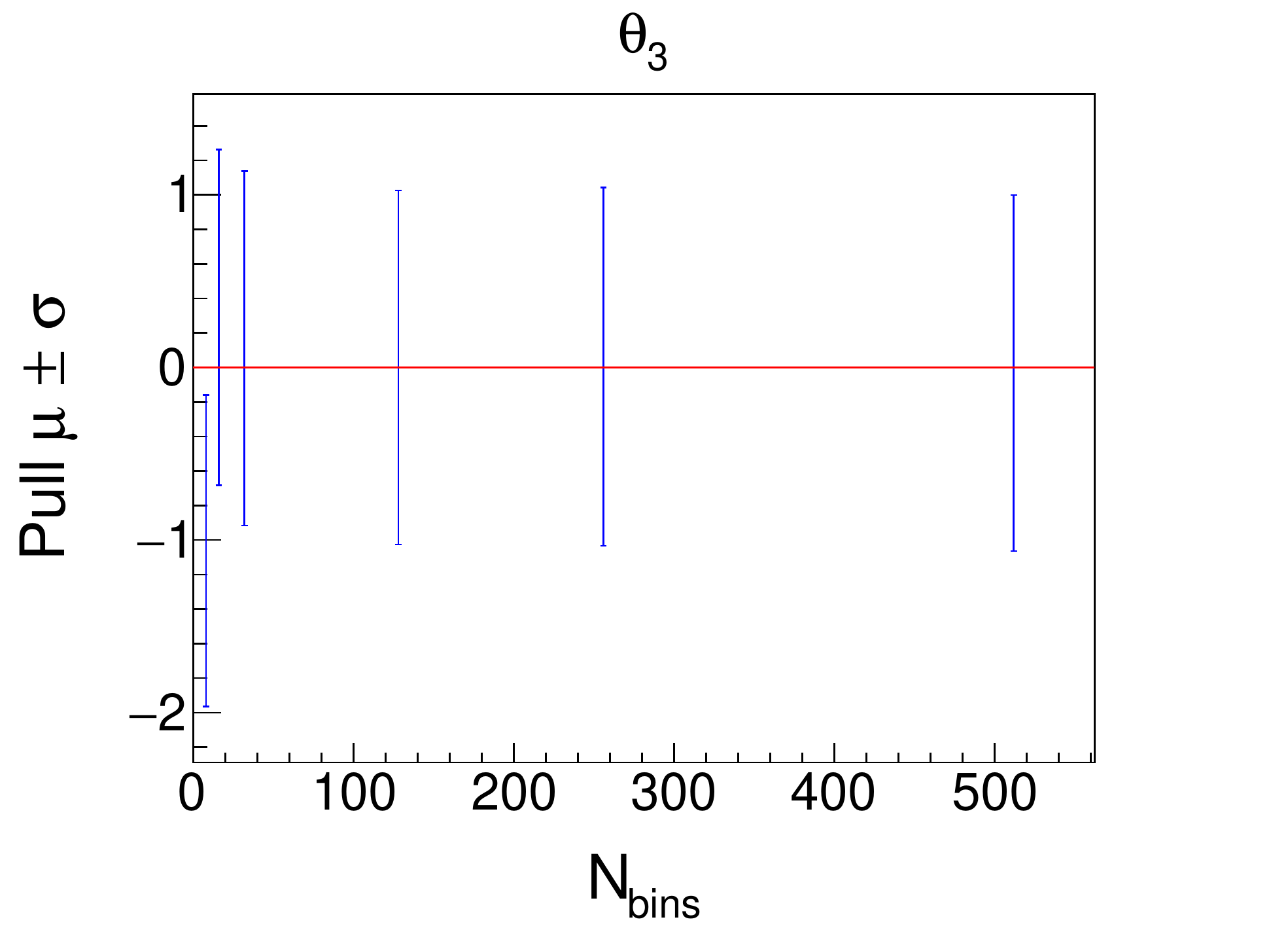}
    \end{subfigure}
    \begin{subfigure}[b]{0.39\textwidth}
        \centering
        \includegraphics[width=\textwidth]{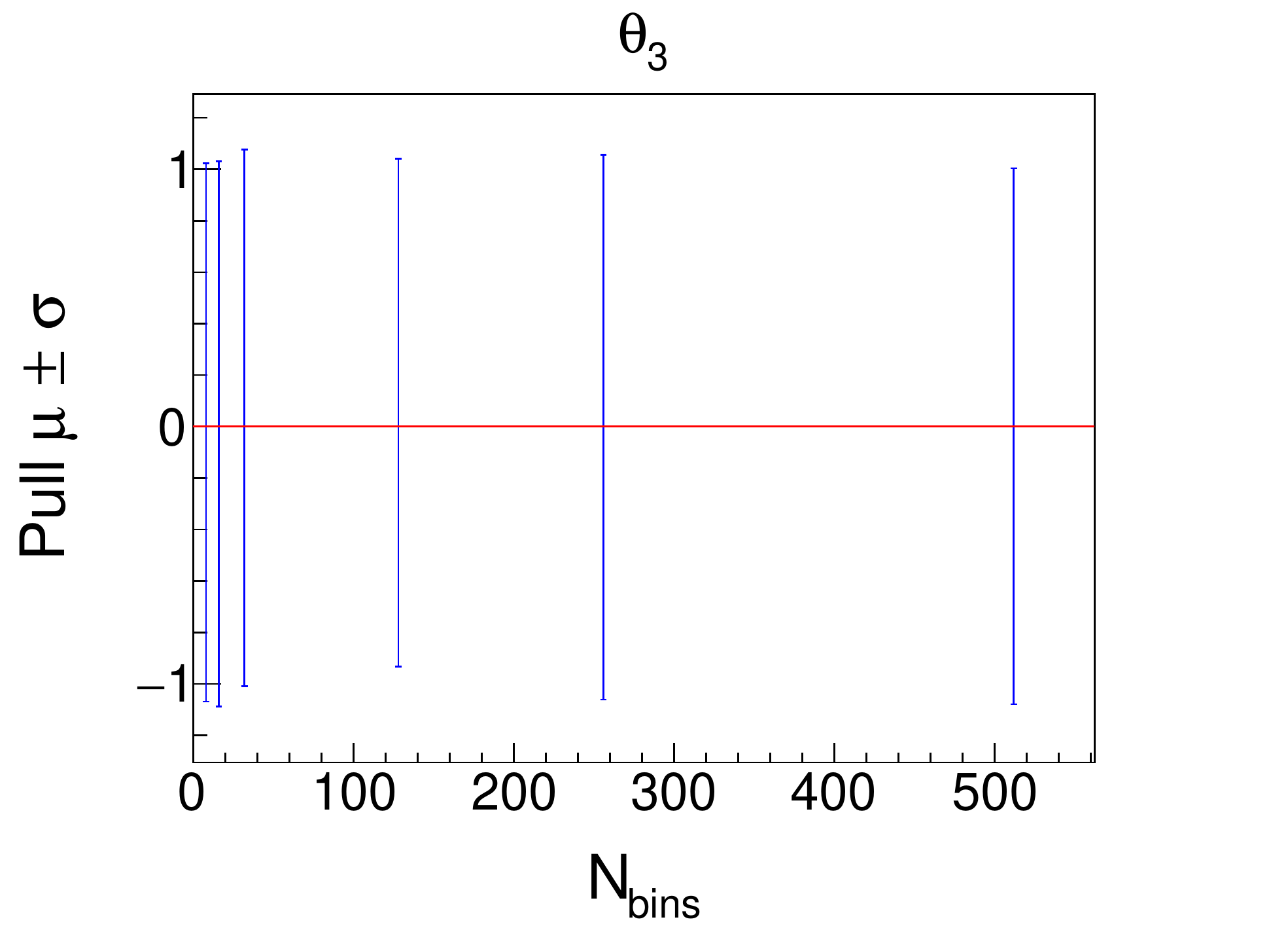}
    \end{subfigure}
    \begin{subfigure}[b]{0.39\textwidth}
        \centering
        \includegraphics[width=\textwidth]{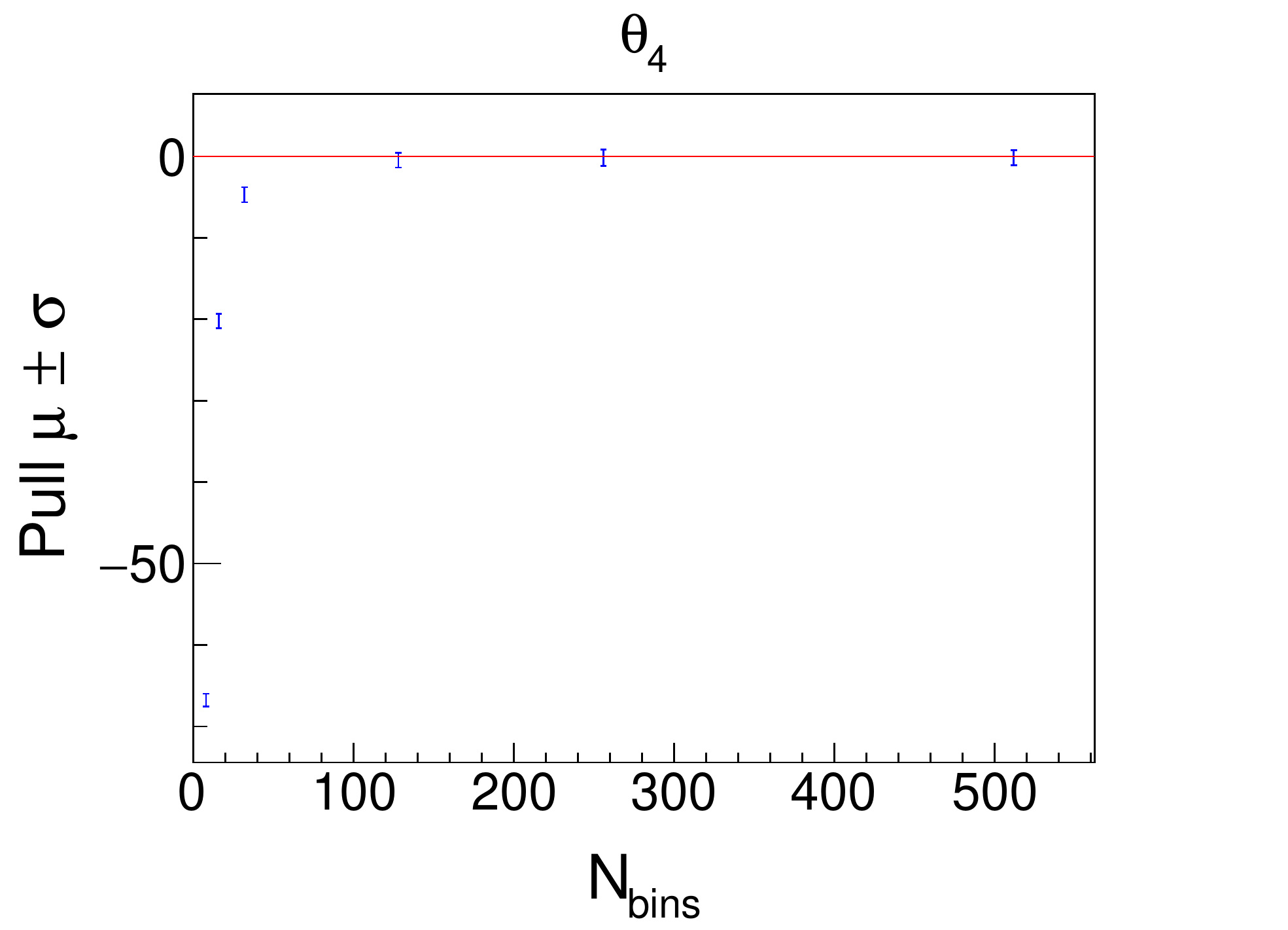}
    \end{subfigure}
    \begin{subfigure}[b]{0.39\textwidth}
        \centering
        \includegraphics[width=\textwidth]{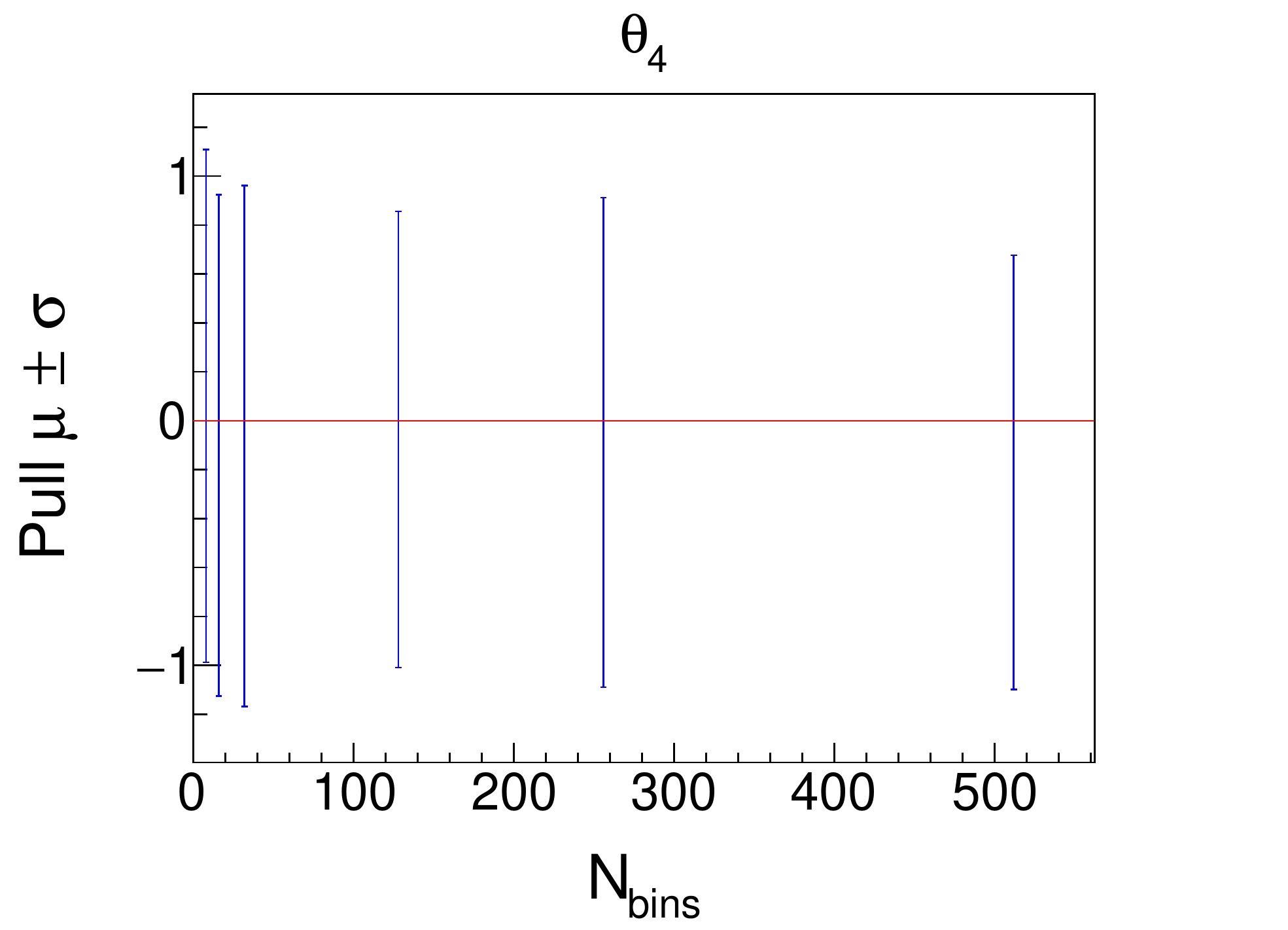}
    \end{subfigure}
    \caption{Mean and standard deviation of the pull as a function of \Nbins from a Gaussian fit for various fit parameters with respect to the generated and fitted values. Fits are performed with the default (left) and \textsc{RooBinSamplingPdf} (right) methods. A thousand data sets are generated and fitted for each value of \Nbins.}
    \label{fig:ATLAS-pulls-Nbins}
\end{figure}

\begin{figure}[hbtp]
    \centering
    \begin{subfigure}[b]{0.39\textwidth}
        \centering
        \includegraphics[width=\textwidth]{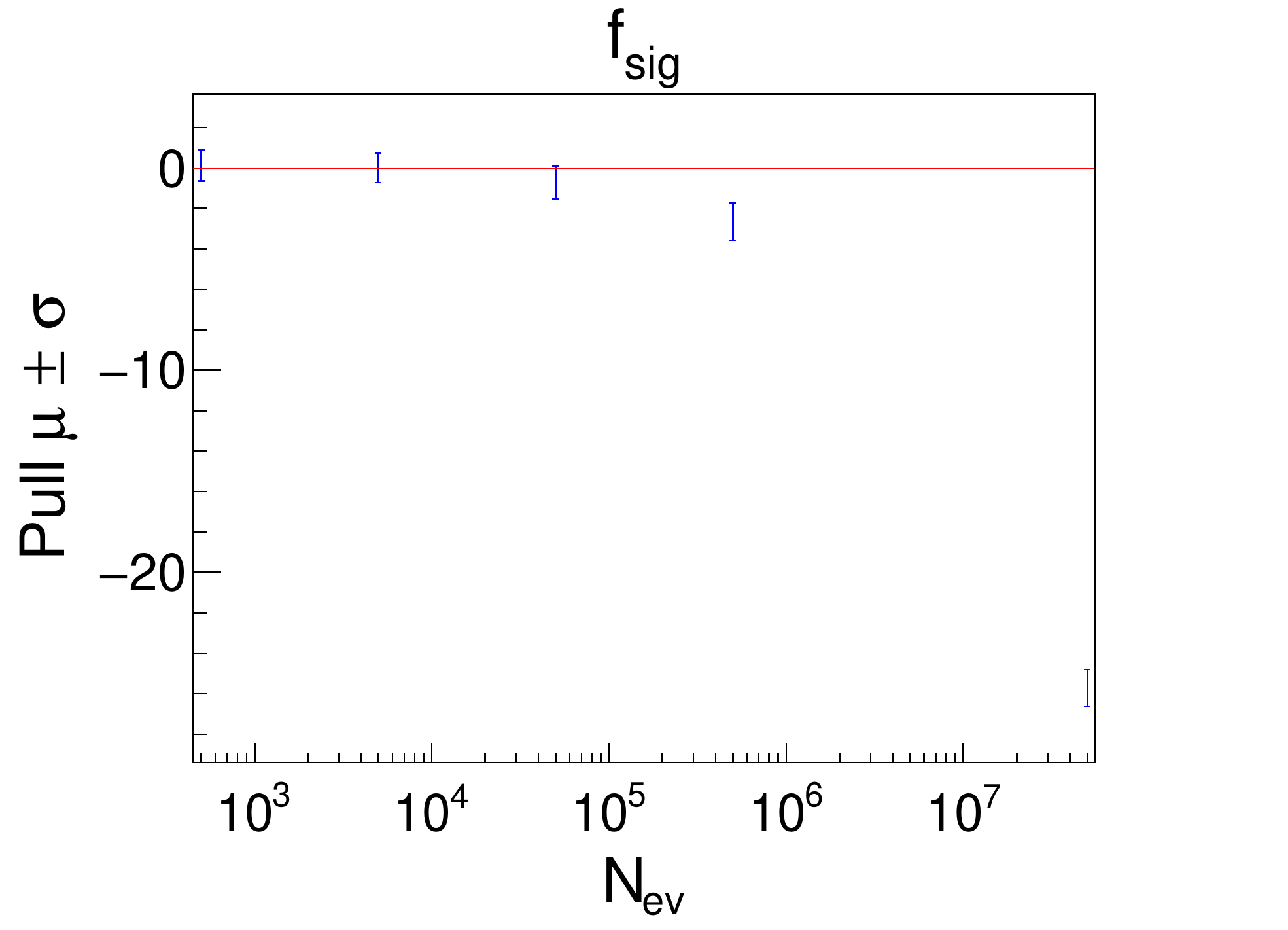}
    \end{subfigure}
    \begin{subfigure}[b]{0.39\textwidth}
        \centering
        \includegraphics[width=\textwidth]{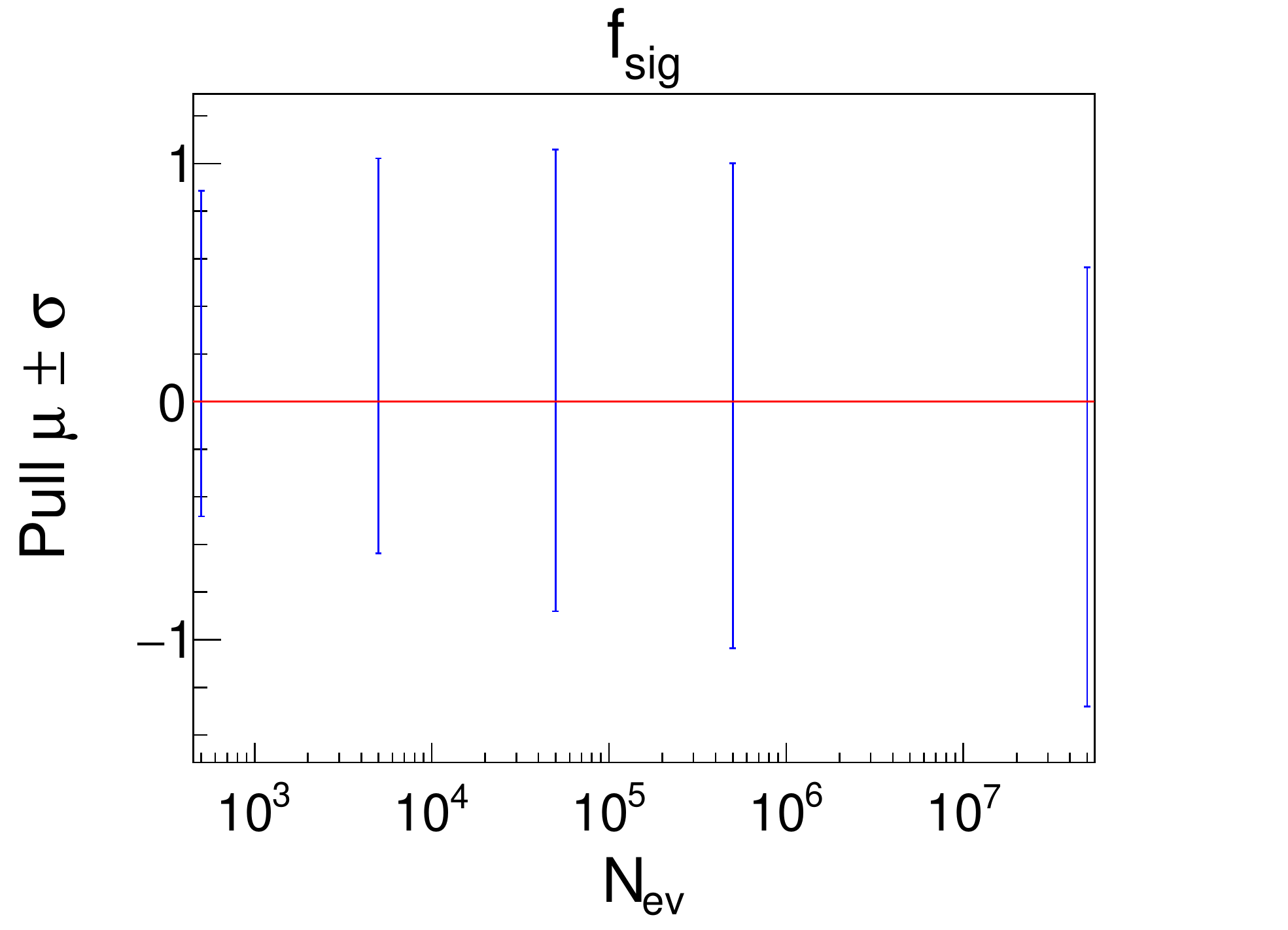}
    \end{subfigure}\\
    \begin{subfigure}[b]{0.39\textwidth}
        \centering
        \includegraphics[width=\textwidth]{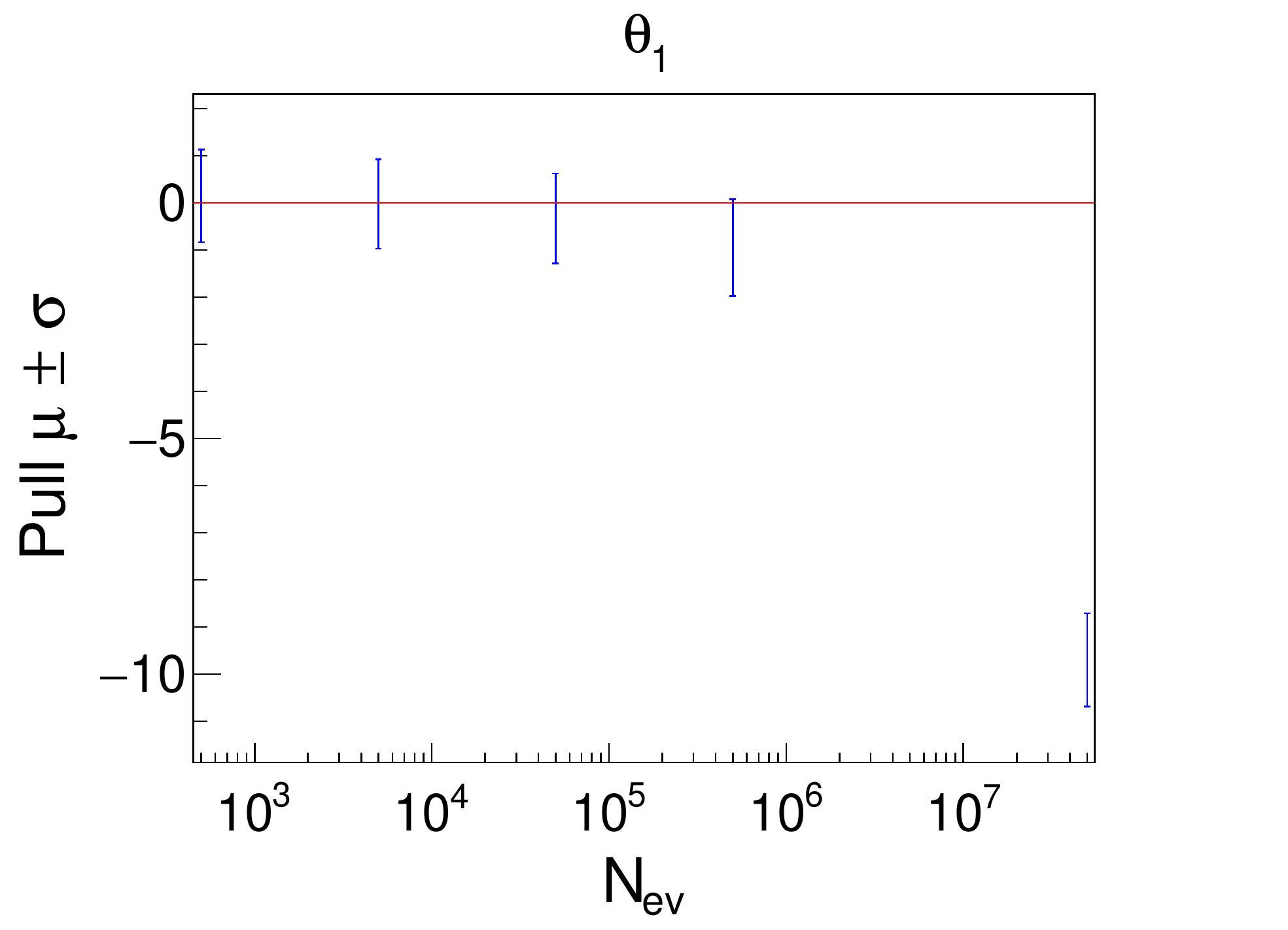}
    \end{subfigure}
    \begin{subfigure}[b]{0.39\textwidth}
        \centering
        \includegraphics[width=\textwidth]{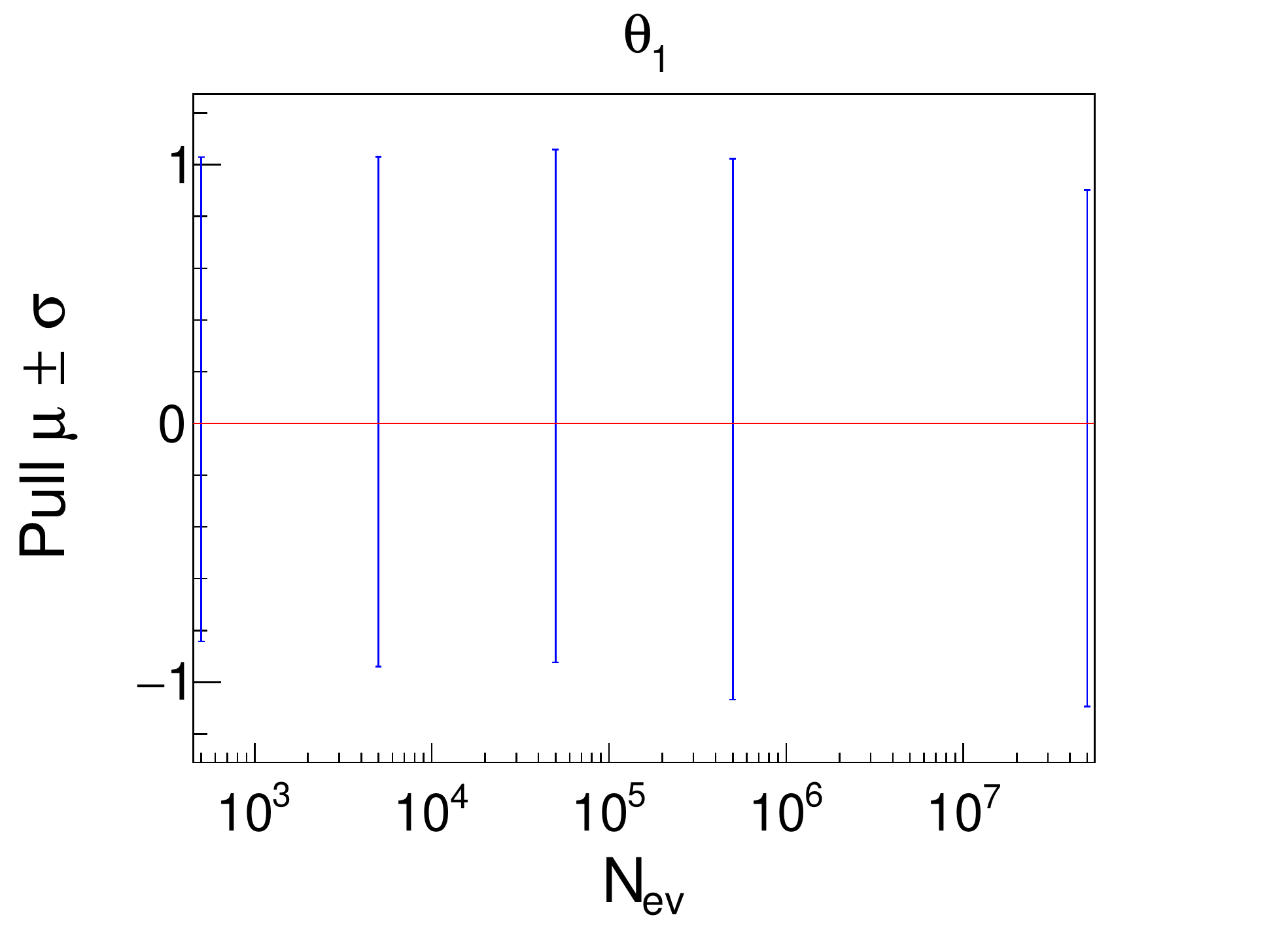}
    \end{subfigure}
    \begin{subfigure}[b]{0.39\textwidth}
        \centering
        \includegraphics[width=\textwidth]{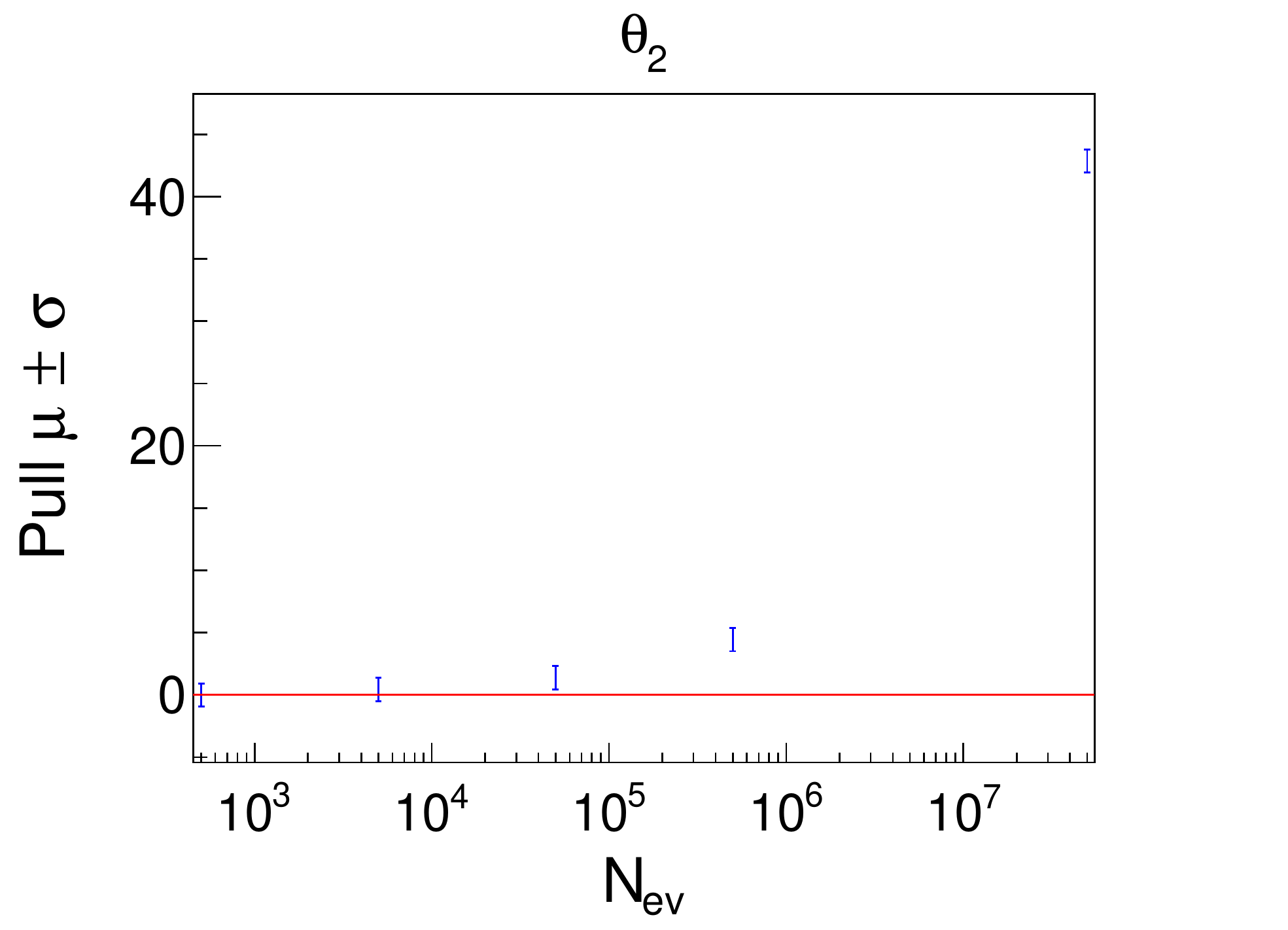}
    \end{subfigure}
    \begin{subfigure}[b]{0.39\textwidth}
        \centering
        \includegraphics[width=\textwidth]{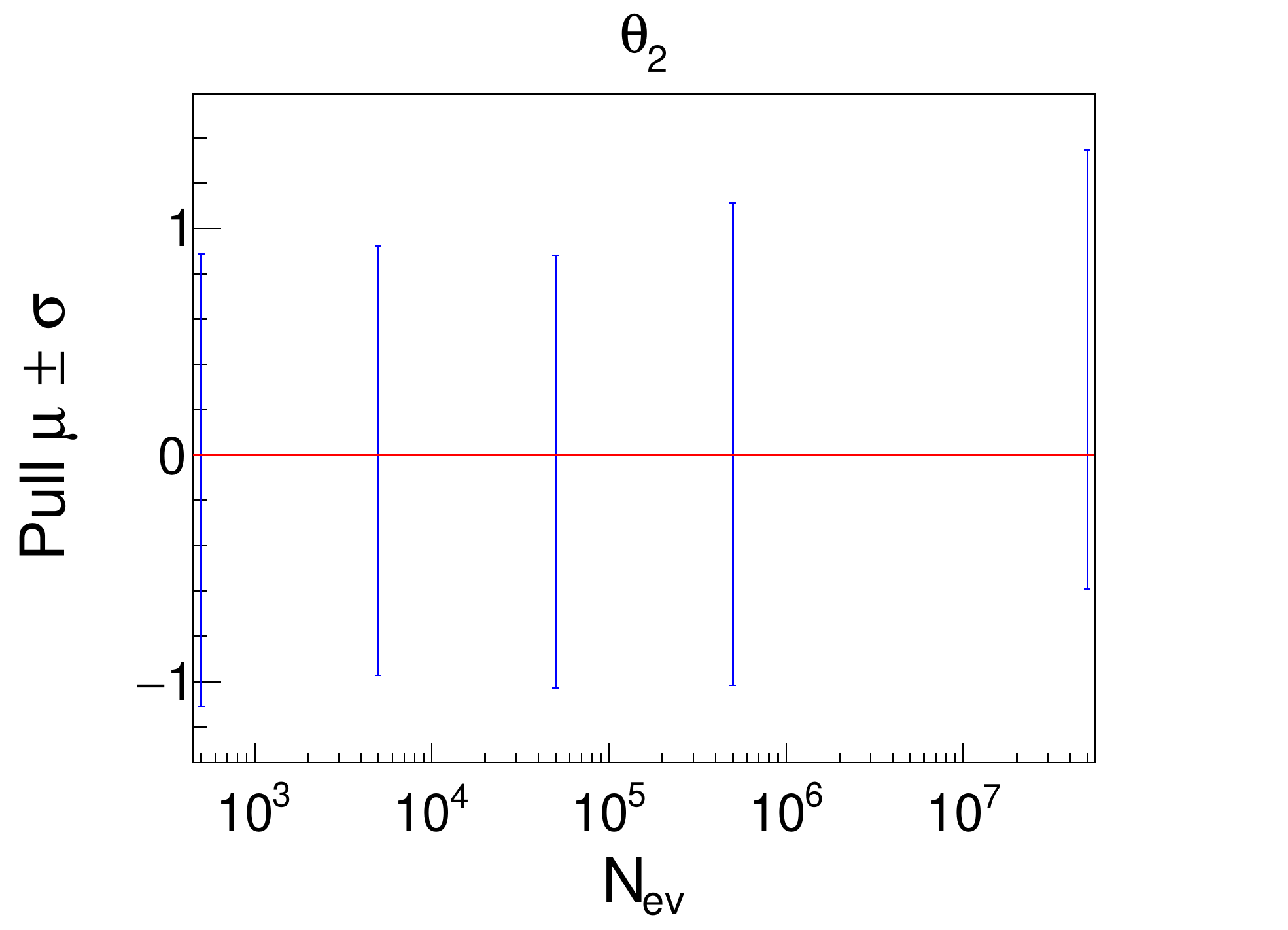}
    \end{subfigure}
    \begin{subfigure}[b]{0.39\textwidth}
        \centering
        \includegraphics[width=\textwidth]{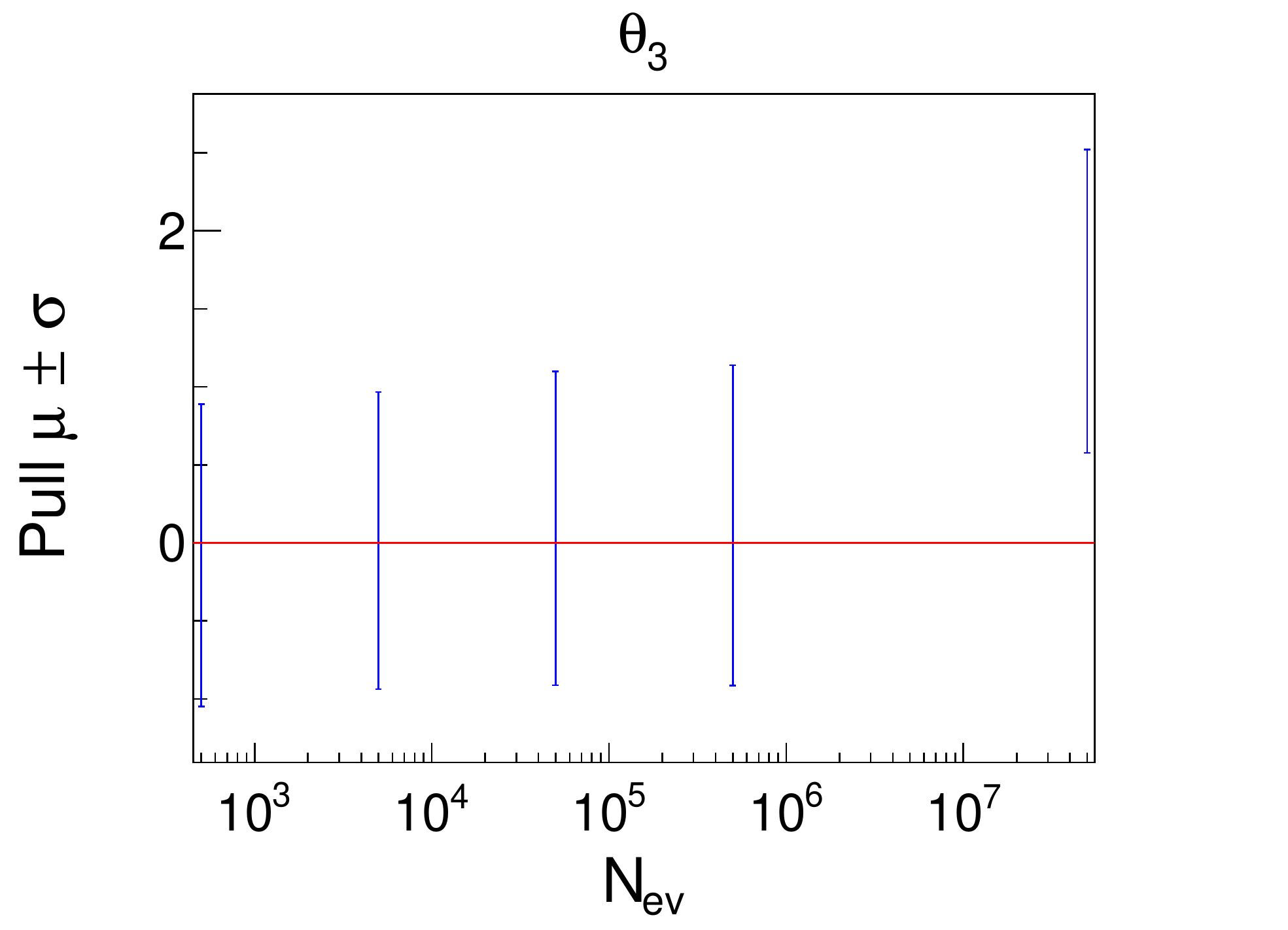}
    \end{subfigure}
    \begin{subfigure}[b]{0.39\textwidth}
        \centering
        \includegraphics[width=\textwidth]{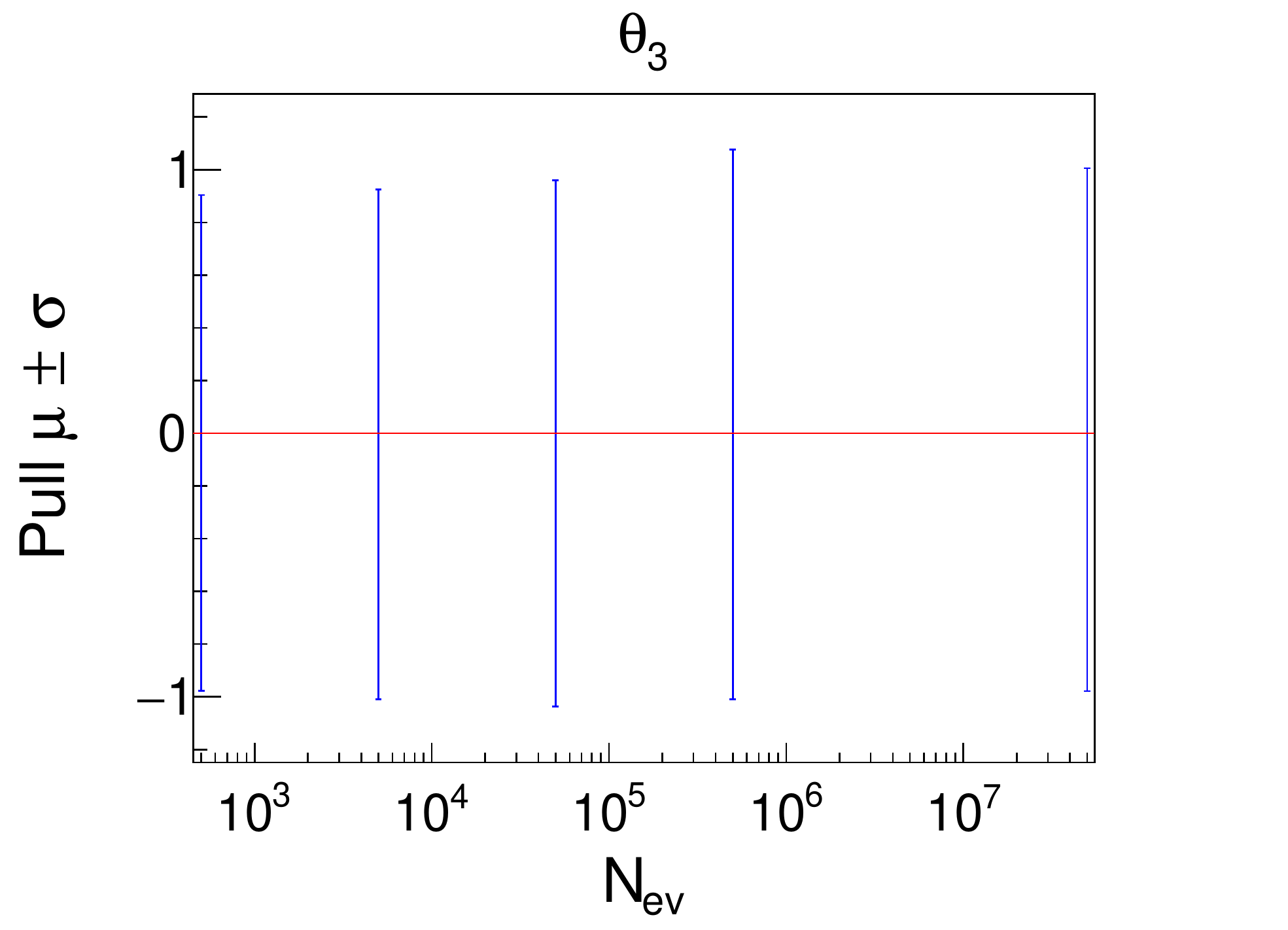}
    \end{subfigure}
    \begin{subfigure}[b]{0.39\textwidth}
        \centering
        \includegraphics[width=\textwidth]{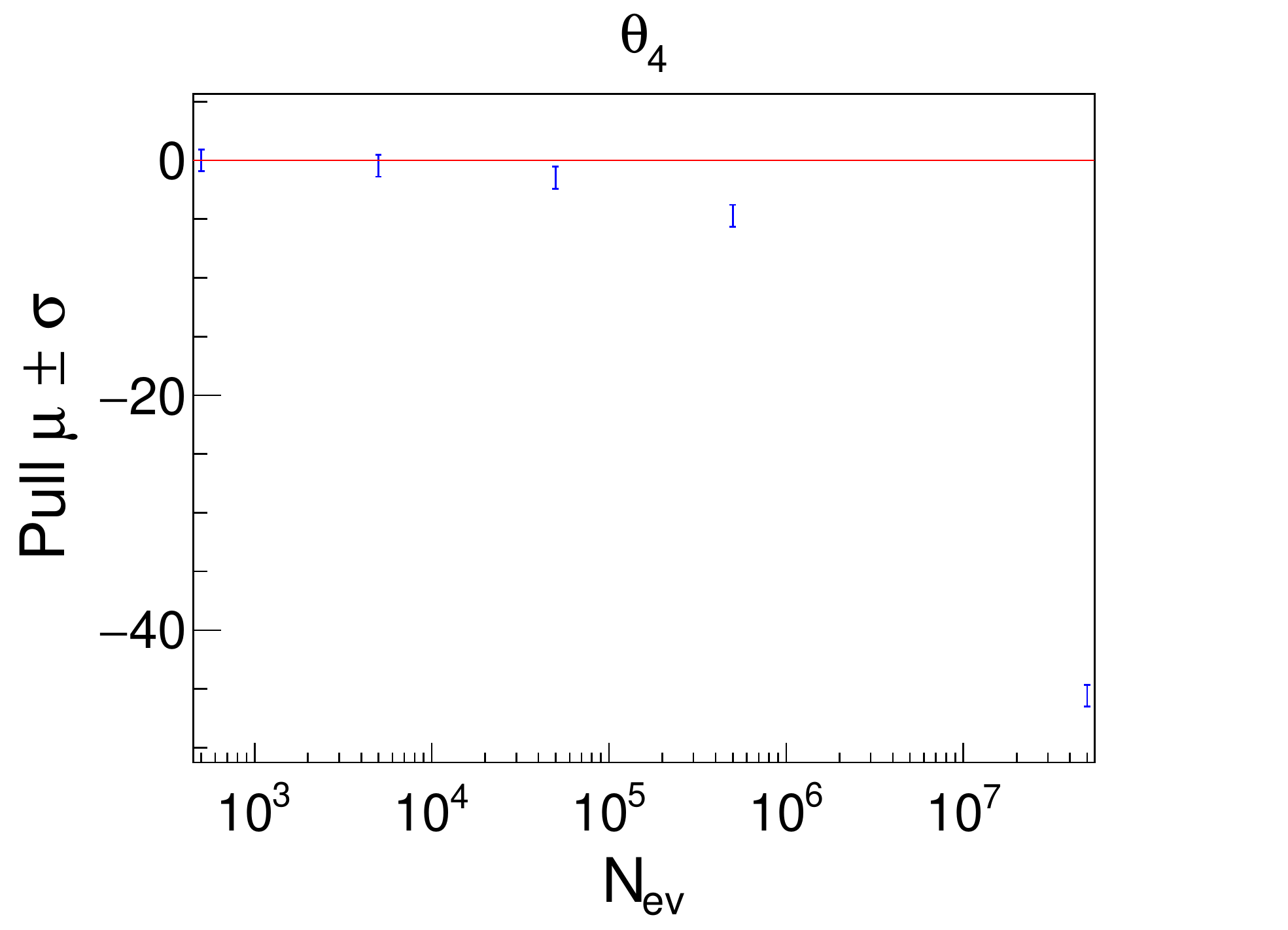}
    \end{subfigure}
    \begin{subfigure}[b]{0.39\textwidth}
        \centering
        \includegraphics[width=\textwidth]{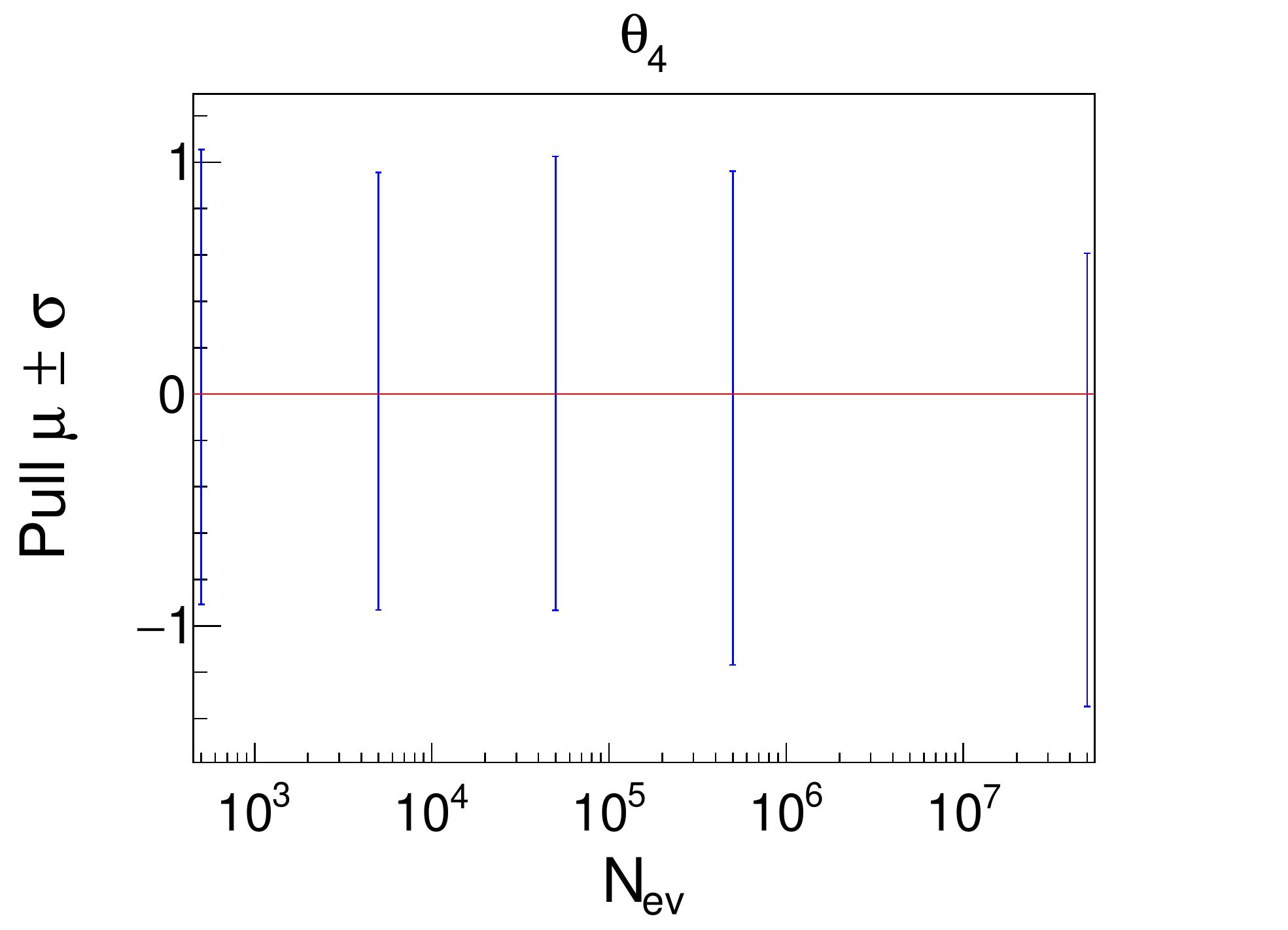}
    \end{subfigure}
    \caption{Mean and standard deviation of the pull as a function of \Nev from a Gaussian fit for various fit parameters with respect to the generated and fitted values. Fits are performed with the default (left) and \textsc{RooBinSamplingPdf} (right) methods. A thousand data sets are generated and fitted for each value of \Nev. The $x$-axis is shown in logarithmic scale.}
    \label{fig:ATLAS-pulls-Nev}
\end{figure}

The pull for the fit parameters with respect to the generated and fitted values, expected to be centred around $\mu=0$ with a standard deviation of $\sigma=1$, is fitted with a Gaussian function.
The dependence of the bias in all of the fit parameters is shown in \cref{fig:ATLAS-pulls-Nbins,fig:ATLAS-pulls-Nev} for six values of \Nbins and five values of \Nev, respectively. 
It is evident that a default binned fit induces significant biases, unless a large \Nbins or a small \Nev is assumed. In fact, expected biases smoothly decrease as a function of the increasing (decreasing) \Nbins(\Nev). Sometimes, as for example looking at \fsig for $\Nbins=16$, the estimate of the fit-parameter standard deviation may be biased, too. 
On the contrary, when exploiting the \textsc{RooBinSamplingPdf}, the estimate of fit-parameter central values and standard deviations appears robust on a large ensemble of \Nbins and \Nev values. Small deviations with very small \Nbins or very large \Nev, see for instance $\Nev= \num{50E6}$ in \cref{fig:ATLAS-pulls-Nev}, and the \textsc{RooBinSamplingPdf} may be recovered by requiring higher integration precision.

\section{Conclusions and future work}
Currently, all PDFs are integrated numerically, even if analytic integrals are known to \roofit.
This can be changed should the need arise.
Furthermore, the method is restricted to one-dimensional PDFs.
Similarly, this can be extended to multi-dimensional PDFs if necessary.

\acknowledgments
TN acknowledges support from the Swiss National Science Foundation through grant 185050. VVG and DYT acknowledge support of the European Research Council under Consolidator grant RECEPT 724777. AS acknowledges support from the U.S. Department of Energy, Office of Science, under grant DE-SC0010107.

\FloatBarrier
\bibliographystyle{JHEP}
\bibliography{bibliography}

\end{document}